\documentclass[10pt,journal,compsoc]{IEEEtran}
\usepackage{graphicx}
\usepackage{textcomp}
\usepackage{xcolor}
\usepackage{threeparttable}
\usepackage{multirow}
\usepackage{subfigure}
\usepackage{threeparttable}
\usepackage{caption}
\usepackage{amsthm}
\usepackage{epstopdf}
\usepackage{enumitem}
\usepackage{cite}
\usepackage{amsmath,amssymb,amsfonts}
\usepackage{algorithm,algorithmic}
\usepackage{url}
\newtheorem{definition}{Definition}
\def\BibTeX{{\rm B\kern-.05em{\sc i\kern-.025em b}\kern-.08em
		T\kern-.1667em\lower.7ex\hbox{E}\kern-.125emX}}
\begin{document}
\title{Spatial Crowdsourcing Task Allocation Scheme for Massive Data with Spatial Heterogeneity}

\author{Kun Li, Shengling Wang*, ~\IEEEmembership{Senior Member,~IEEE}, Hongwei Shi, Xiuzhen Cheng, ~\IEEEmembership{Fellow,~IEEE}, Minghui Xu
	\IEEEcompsocitemizethanks{
		\IEEEcompsocthanksitem Kun Li is with School of Artificial Intelligence, Beijing Normal University (BNU), and also with the School of Computer Science and Technology, Shandong University, Qingdao 266510, China, E-mail: likun@mail.bnu.edu.cn.
		\IEEEcompsocthanksitem Shengling Wang (Corresponding author) and Hongwei Shi are with School of Artificial Intelligence, Beijing Normal University (BNU), China. Email: wangshengling@bnu.edu.cn, hongweishi@mail.bnu.edu.cn.
		\IEEEcompsocthanksitem Xiuzhen Cheng and Minghui Xu are with the School of Computer Science and Technology, Shandong University, Qingdao 266510, China. E-mail: xzcheng@sdu.edu.cn, mhxu@sdu.edu.cn.}
	\thanks{This work has been supported by the National Natural Science Foundation of China (No.61772044, 62077044,62293555, 62232010, 62302266), the Major Program of Science and Technology Innovation2030 of China(No.2022ZD0117105), Shandong Science Fund for Excellent Young Scholars (No.2023HWYQ-008), Shandong Science Fund for Key Fundamental Research Project (ZR2022ZD02), and the Fundamental Research Funds for the Central Universities.}}
\maketitle
\begin{abstract}
	Spatial crowdsourcing (SC) engages large worker pools for location-based tasks, attracting growing research interest. However, prior SC task allocation approaches exhibit limitations in computational efficiency, balanced matching, and participation incentives. To address these challenges, we propose a graph-based allocation framework optimized for massive heterogeneous spatial data. The framework first clusters similar tasks and workers separately to reduce allocation scale. Next, it constructs novel non-crossing graph structures to model balanced adjacencies between unevenly distributed tasks and workers. Based on the graphs, a bidirectional worker-task matching scheme is designed to produce allocations optimized for mutual interests. Extensive experiments on real-world datasets analyze the performance under various parameter settings.
\end{abstract}
\begin{IEEEkeywords}
Spatial Crowdsourcing, Tasks Allocation, Non-crossing Graph.
\end{IEEEkeywords}

\section{Introduction}\label{sec:intro}
Mobile crowdsourcing has emerged as an Internet-enabled paradigm for distributed problem-solving, harnessing collective intelligence to reduce costs and improve efficiency. The proliferation of location-aware applications has led to widespread interest in spatial crowdsourcing (SC) - integrating crowdsourcing with geolocation information. SC has been utilized in diverse domains including data collection\cite{chen2014gmission}, traffic planning\cite{kong2018lotad}, online ride-hailing services\cite{xu2020privacy}, and disaster monitoring\cite{panteras2018enhancing}.

\begin{figure}[htbp]
	\centering
	\includegraphics[width=0.4\textwidth]{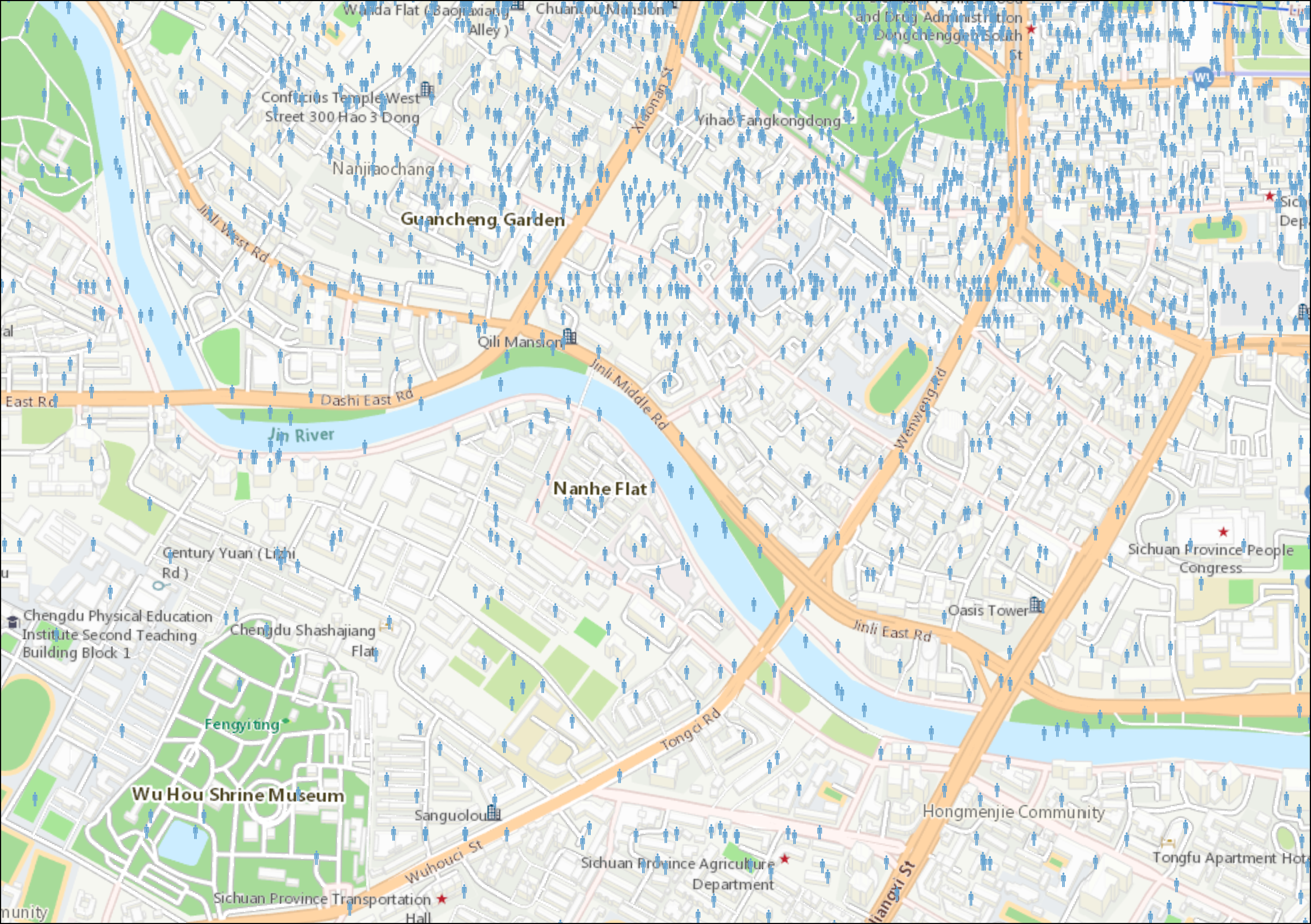}
	\caption{The distribution of online car-hailing orders near Jin River watershed in Chengdu City.}
	\label{fig:order}
\end{figure}

\begin{figure*}
	\centering  
	\subfigbottomskip=1pt 
	\subfigcapskip=-5pt 
	\subfigure[Balanced distribution of tasks and workers.]{
		\label{fig:poi1}
		\includegraphics[width=0.3\linewidth]{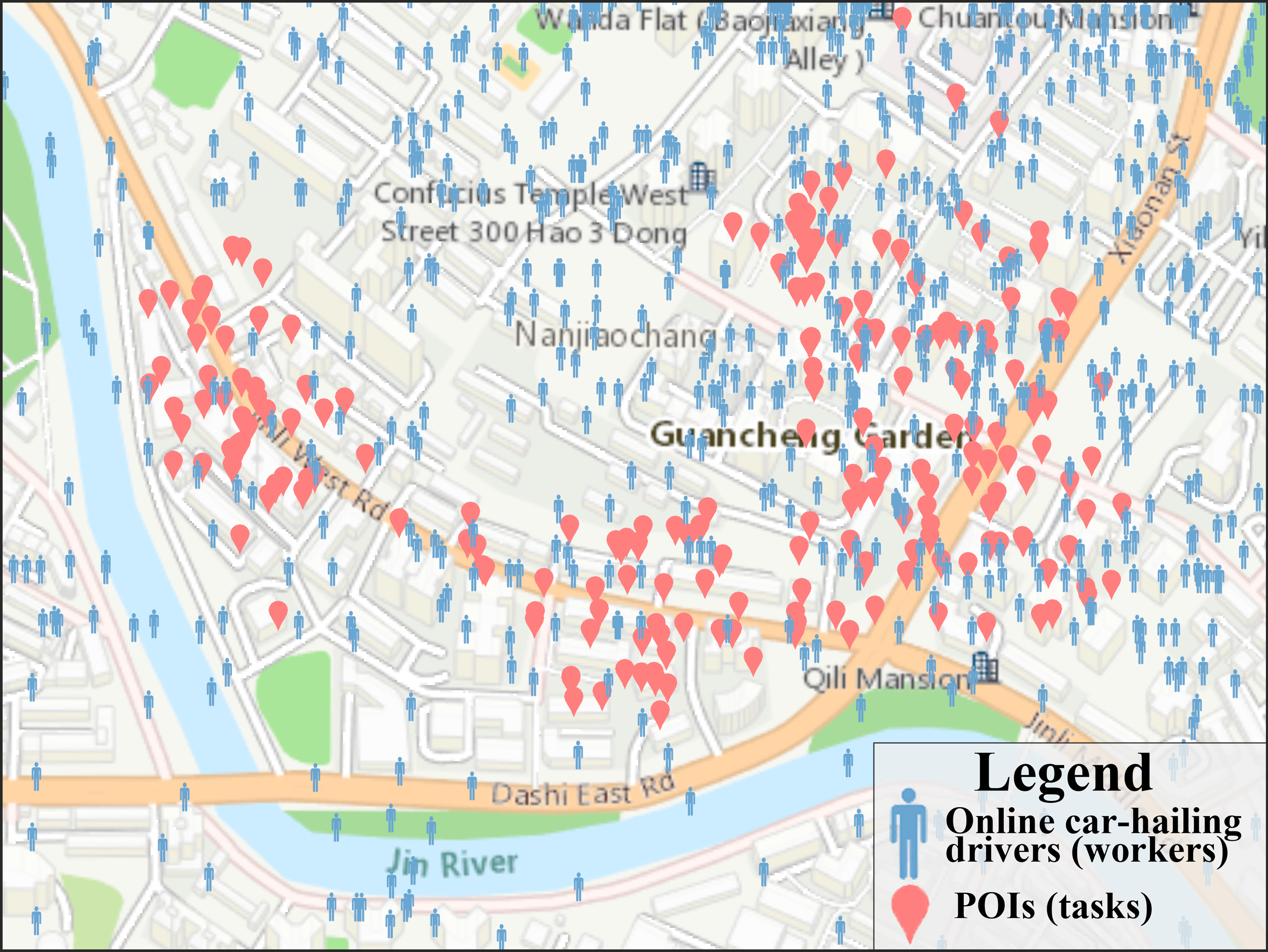}}
	\subfigure[Uneven distribution with few tasks and many workers.]{
		\label{fig:poi2}
		\includegraphics[width=0.3\linewidth]{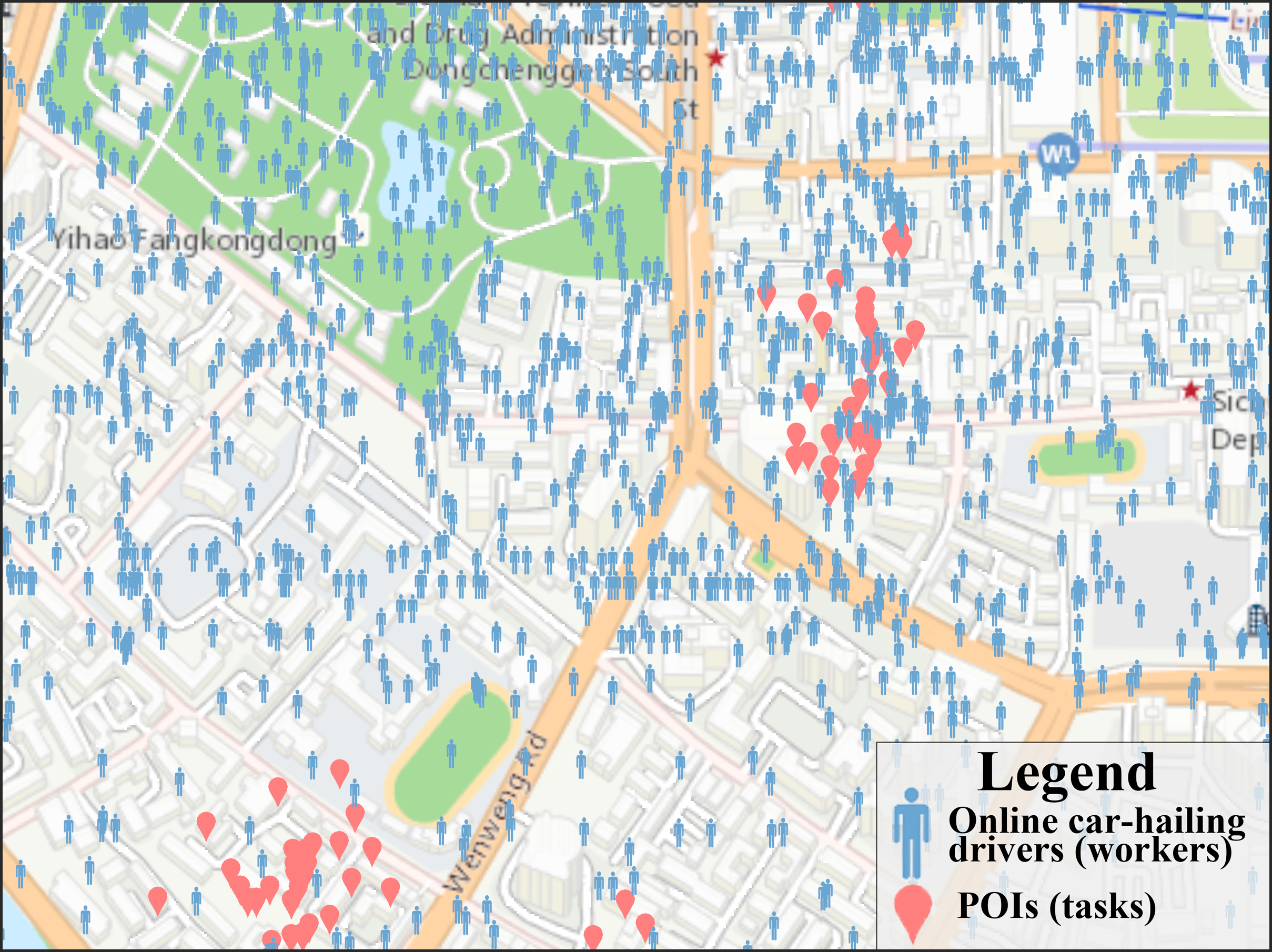}}
	\subfigure[Uneven distribution with many tasks and few workers.]{
		\label{fig:poi3}
		\includegraphics[width=0.3\linewidth]{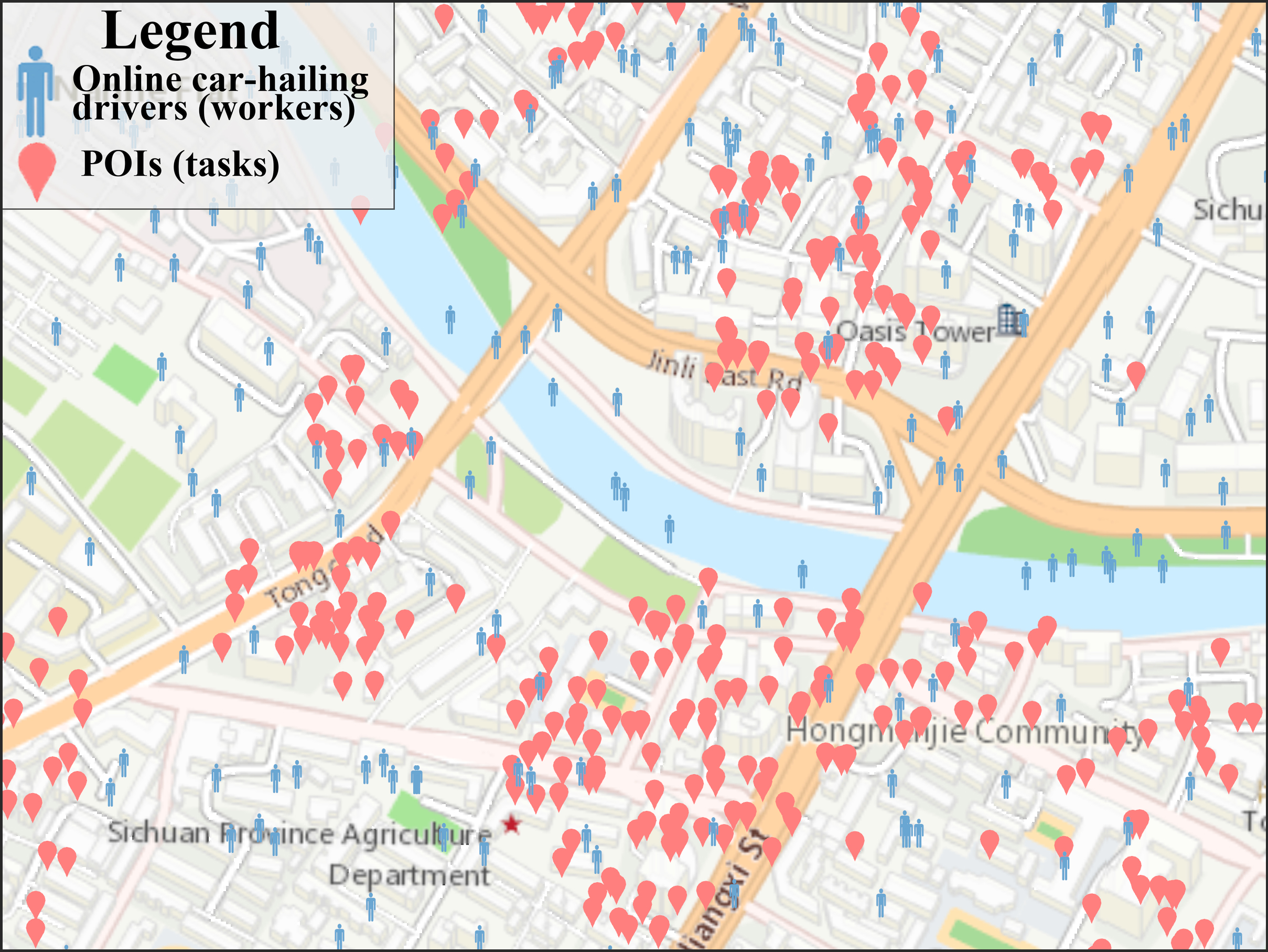}}		
	\caption{Three distributions of tasks and workers in online car-hailing scenario near Jin River watershed in Chengdu City.}		
	\label{fig:poi}
\end{figure*}

Designing an efficient task allocation algorithm is crucial for spatial crowdsourcing, given its significant impact on system effectiveness and reliability. However, high-dimensional heterogeneous data and complex task requirements impose challenges for allocation schemes. Most prior studies focus on feature engineering of workers and tasks to enable allocation via classification and prediction. While considering incentives and budgets, these approaches exhibit three core limitations:
\begin{itemize}
	\item Inefficient task allocation due to intensive computation. Real-world spatial crowdsourcing involves massive datasets - for instance, 70 million online ride-hailing orders were generated in Chengdu in one month \cite{gaia}, as exemplified in Fig. \ref{fig:order}. Such scale incurs prohibitive computation costs, hindering system efficiency.
	\item Allocation imbalance due to supply-demand mismatches. Spatial heterogeneity creates uneven worker and task densities, such as dense tasks but few workers in some areas, and excessive workers but few tasks in others as shown in Fig. \ref{fig:poi}. Distance-based filtering exacerbates such imbalance issues, reducing efficiency.
	\item Weak incentive for SC participation due to one-sided benefit maximization. The advantage of SC is to use the competition and cooperation among a large number of participants recruited through the Internet to achieve a higher quality of service. Win-win has always been one of the goals pursued in task allocation, and pursuing only worker or requester benefit maximization discourages participation and loses crowdsourcing advantages of mutual gains.
\end{itemize}

To address the above challenges, we propose a spatial crowdsourcing task allocation framework comprising downsizing, reconfiguration, and allocation modules. The core contributions are:
\begin{itemize}
	\item Improved efficiency via batch allocation. The downsizing module clusters similar tasks/workers enabling batch allocation, superior to one-to-one allocation.
	\item Balanced supply-demand matching through graph reconfiguration. The reconfiguration module establishes novel graph structures to improve the heterogeneity in spatial distributions between tasks and workers, avoiding imbalanced allocation.
    \item Mutual benefits via bidirectional matching. The allocation module incorporates both worker and requester interests, achieving win-win allocation through bidirectional matching.
\end{itemize}

The remainder of this paper is organized as follows. Section \ref{sec:related} reviews prior work on task allocation for spatial crowdsourcing. Section \ref{sec:overview} presents an overview of our proposed allocation framework, with details provided in Sections \ref{sec:downsizing}, \ref{sec:reconfiguration}, and \ref{sec:allocation} on the downsizing, reconfiguration, and allocation modules respectively. Section \ref{sec:evaluation} describes our experimental evaluation on real-world data to assess the performance of our approach. Finally, Section \ref{sec:conclusion} concludes the paper.

\section{Related work}\label{sec:related}
According to \cite{reddy2010recruitment}, it is important to select the appropriate workers to perform the SC tasks as it directly affects the quality of the crowdsourcing results. In recent years, there have been many research works on task allocation schemes in SC scenarios. In general, we classify the existing works into two categories: feature-based and constraint-based.

Feature-based task allocation schemes usually focus on the degree of fitness of workers and tasks in various aspects. Such schemes tend to first identify the important features in crowdsourcing, such as task categories, required competencies, or workers' behavioral patterns and preferences, and then use various algorithms to extract the features of all workers and tasks, and finally allocate tasks to workers with the same or similar features. For example, in \cite{wang2020towards} Wang Z \emph{et al.} designed a worker selection mechanism based on a detailed representation of worker characteristics. They make full use of a variety of information such as task category, task description, task time, and task location, combined with invisible feedback from worker attendance, to accurately model worker preferences for tasks. With the rapid development of machine learning and deep learning algorithms, related methods have been introduced to crowdsourcing. For fog-based crowdsourcing applications in IoT networks, Yu Y \emph{et al.} in \cite{yu2019reliable} developed a spatiotemporal attribute learning model based on user behavior, which uses the user's interest attribute matching model to identify the candidate nodes that meet the requirements of crowdsourcing tasks. In \cite{ye2021task}, Ye G \emph{et al.} proposed a two-stage task allocation framework based on geographic partitioning in SC scenarios. They use reinforcement learning methods in which a graph neural network with the attention mechanism is used to learn the embeddings of allocation centers, delivery points, and workers. In contrast to the feature extraction methods mentioned above, Shi Z \emph{et al.} in \cite{shi2021crowdsourcing} creatively proposed a Bayesian probabilistic model called Gaussian Latent Topic Model (GLTM). By using GLTM, the authors propose a truth value inference algorithm and use it to accurately infer the truth value and topic of a task, and dynamically update the topic-level reliability of workers, which can be seen as the characteristics of tasks and workers.

Constraint-based task allocation schemes tend to be more concerned with the external constraints of task allocation, such as allocation rate, task budget, expected quality, and worker utility. Such methods tend to construct corresponding mathematical models for different scenarios, and then compute the optimal solution of the objective function under the constraints to obtain the optimal task allocation scheme, and further analyze the effects of different parameters on the results. Most of the research works take the budget as the constraint. In \cite{wu2018toward}, Wu P \emph{et al.} proposed a real-time budget-constrained SC task allocation scheme, which aims to maximize both the task allocation rate and the expected quality of the results with a limited budget. Wang L \emph{et al.}, on the other hand, focus on constraints such as spatial constraints, effective duration, operational complexity, budget constraints, and the number of workers required for tasks in the SC scenario, and work to find a task allocation scheme that maximizes task coverage and data quality under the budget constraints\cite{wang2017multi, wang2018heterogeneous}. They abstract the problem as a mathematical problem of finding the Pareto optimal allocation scheme for a multiobjective optimization problem under the condition of minimizing the incentive budget. Liang D \emph{et al.} focus more on the phenomenon of competition and cooperation in the task allocation process and use it as a constraint\cite{liang2021novel}. They designed a two-stage three-way decision model consisting of a competition-optimization model and a negotiation-cooperation model.


However, the computational requirements imposed by high-dimensional data remain a core challenge for efficiency in the above task allocation approaches. To mitigate this inefficiency, group-based allocation has gained traction as an emerging technique, assigning tasks to worker groups rather than individuals. By allocating at the group level, computational load can be significantly reduced. Wang W \emph{et al.} in \cite{wang2018strategic} noted the dishonest behavior of workers with social connections in completing crowdsourcing tasks as team collaboration, which is detrimental to other workers and requesters. Their proposed mechanism selects a fraction of workers to form an optimal team based on their social structure, ability, and cost. In contrast to the artificial group formation scheme described above, Jiang J \emph{et al.} in \cite{jiang2019group} discusses the task allocation scheme for naturally existing groups of workers. They proposed the concept of contextual crowdsourcing value to measure the ability of a natural group to complete a crowdsourcing task by coordinating with other groups, and then designed a group-oriented task allocation scheme based on it.


In real-world systems, task allocation is often conducted locally rather than globally - candidates are filtered by metrics like location before matching. Distance-based filtering is a prevalent approach. However, existing methods overlook the spatial heterogeneity inherent to crowdsourcing, and distance-based filtering exacerbates the allocation imbalance caused by uneven distributions. To address this, we propose an allocation scheme tailored for uneven distributions. Further, incorporating both worker and requester interests synthesizes perspectives to motivate participation.

\section{Overview of our framework}\label{sec:overview}
The task allocation system designed to solve the problems mentioned in Section \ref{sec:intro} can be divided into the following modules:
\begin{enumerate}
	\item Downsizing Module: This module clusters tasks and workers separately. Doing so enables batch matching between task and worker clusters. This allows efficient allocation of proximate tasks to similar workers and bundling of tasks for individuals. It improves efficiency and reduces computational load.
	\item Reconfiguration Module: In this module, we model task clusters and worker clusters as nodes, and construct edges between them based on their locations and regional node density. Performing allocation using these graph-based adjacencies avoids problems like spatial heterogeneity arising from distance-based screening, as described in Section \ref{sec:intro}.
	\item Allocation Module: This module contains submodules for evaluation and matching. It assesses task and worker clusters separately to establish allocation priorities. Then it performs bi-directional allocation between prioritized clusters. This integrated approach optimizes allocations for worker utility, requester revenue, and allocation rate.
\end{enumerate}

\begin{figure*}[htbp]
	\centering
	\includegraphics[width=0.9\textwidth]{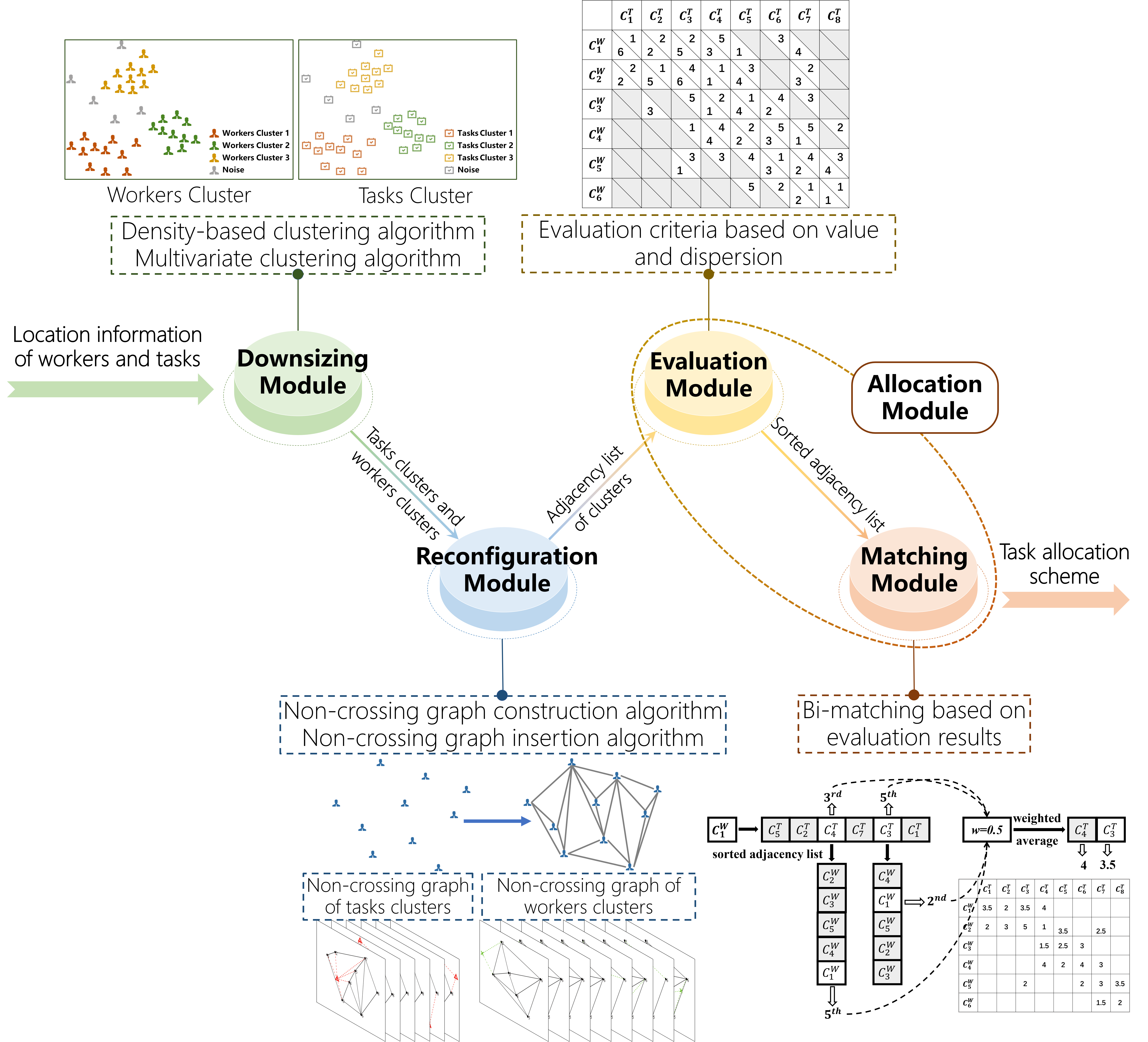}
	\caption{Overview of our framework.}
	\label{fig:framework}
\end{figure*}

Our proposed spatial crowdsourcing task allocation approach comprises the following steps (Fig. \ref{fig:framework}):
\begin{enumerate}
	\item Independently cluster tasks and workers based on their attributes.
	\item Model task and worker clusters as nodes to construct graph adjacencies, forming novel non-crossing graphs. This establishes adjacency relationships between tasks and workers.
	\item Derive adjacent cluster lists for each task and worker cluster from the graphs.
	\item Evaluate and rank the adjacency lists using defined evaluation functions.
	\item Perform bidirectional matching between task and worker clusters based on the ranked adjacency lists.
	\item Determine the optimal allocation scheme based on the bidirectional matching results.
\end{enumerate}

In the following sections, the modules of the task allocation system are described in detail.

\section{Downsizing module}\label{sec:downsizing}
To improve computational efficiency, we propose a batch matching approach between tasks and workers based on clustering algorithms. When workers receive batched tasks, travel distance is often a key factor. Hence, we cluster tasks spatially to allocate proximate bundles to workers. On the requester side, both worker locations and abilities are important for multi-task assignments. Therefore, our worker clustering integrates ability data and geographic information, enabling batch assignment to workers with similar abilities and locations.

For task clustering, we adopt a density-based algorithm to detect concentrated zones and sparse boundaries between tasks. This approach groups proximate tasks into bundles amenable for batch assignment. For workers, we employ a multivariate clustering technique based on the K-means algorithm, using both location and skill data. This divides workers into clusters that minimize within-cluster differences in terms of abilities and locations. As extensive research already exists on both clustering techniques, we directly apply the well-established unsupervised learning algorithms for task and worker clustering, without repeatedly analyzing the models themselves. The clustering outputs are then utilized in our downstream reconfiguration and matching modules.



Based on the above algorithms, we define the task cluster and worker cluster as follows:
\begin{definition}[Task Cluster]
	A task cluster, denoted as $C^T(t_1,...,t_i,...)$, is defined as a set of tasks with similar characteristics, where $t_i$ represents the $i^{th}$ task.
\end{definition}

\begin{definition}[Worker Cluster]
	A worker cluster, denoted by $C^W(w_1,...,w_i,...)$, is defined as a set of workers with similar characteristics, where $w_i$ represents the $i^{th}$ worker.
\end{definition}

For computational efficiency, we represent each cluster by the coordinate at the center of the minimum enclosing circle covering all the nodes within that cluster.


\section{Reconfiguration module}\label{sec:reconfiguration}
To address the issue of unbalanced matching, we propose an adaptive method to construct adjacencies between worker clusters and task clusters based on the density of their locality. However, building new adjacency relations solely from spatial data is challenging, as complexity factors like social ties and user preferences are not incorporated. To overcome these limitations, we introduce non-crossing graphs - a novel graph model capable of adaptively adjusting edge density according to vertex distribution. By modeling workers and tasks as vertices, and their relations as edges, non-crossing graphs allow optimal adjacency construction without edge crossing. The key advantage is the ability to balance connections and produce an even topology by adaptive edge density control. Our approach thereby provides a robust graph-based solution to balance spatial matching in crowdsourcing systems.

\begin{definition}[Non-crossing graph]
	A non-crossing graph $G=<V,E>$ is an undirected graph where $V=\{v_1,...,v_n\}$ is the set of $n$ vertices and $E=\{e_1,...,e_m\}$ is the set of $m$ edges in $G$. For any two distinct edges $e_1(v_1,v_2)$, $e_2(v_3,v_4)\in E$, if $e_1$ and $e_2$ do not share common endpoints ($v_1\neq v_3$, $v_1\neq v_4$, $v_2\neq v_3$ and $v_2 \neq v_4$), then $e_1$ and $e_2$ will not intersect at any internal points.
\end{definition}

If two nodes $v_i$ and $v_j$ are connected by an edge $e_k(v_i,v_j)$, we say that $v_i$ and $v_j$ are adjacent to each other. Furthermore, if $v_i$ and $v_j$ are not directly connected, but reachable within $k$ hops, they are defined as $k$-layer adjacent vertices of each other.

To construct non-crossing graphs, we propose Algorithm \ref{alg:algorithm1} based on previous graph algorithms \cite{zhou2011computational,8003481}. It takes as input a set of points $S^P$ sorted by descending $x$-coordinates. First, the algorithm initializes the vertex set $V$, convex hull $\text{Conv}(V)$, vertex set $C^P$, edge set $C^E$, and edge set $E$ using the first 3 points (lines 1-4). Then for each subsequent vertex $P_i$, it checks segments $\overline{P_iP_k}$ between $P_i$ and each existing hull vertex $P_k$ in $C^P$. If $\overline{P_iP_k}$ intersects $C^E$ only at $P_k$, then $P_k$ is an adjacent vertex to $P_i$ and $\overline{P_iP_k}$ is added to $E$ (lines 6-11). After each $P_i$, $V$, $\text{Conv}(V)$, $C^P$ and $C^E$ are updated (lines 12-14). This iteratively finds adjacent vertices and adds edges until all vertices are processed. Finally, the non-crossing graph $G=<V,E>$ is output (line 16).
  
\begin{algorithm}[t]
	\caption{Non-Crossing Graph Construction}
	\begin{algorithmic}[1]
		\renewcommand{\algorithmicrequire}{\textbf{Input:}}
		\renewcommand{\algorithmicensure}{\textbf{Output:}}
		\label{alg:algorithm1}
		\REQUIRE $S^P={P_1(x_1,y_1),\ldots,P_n(x_n,y_n)}$: Set of $n$ points sorted by descending $x$-coordinate
		\ENSURE Set of edges representing constructed non-crossing graph
		\STATE $V \gets \{P_1, P_2, P_3\}$;
		\STATE $C^P \gets \{P_i|P_i$ is a vertex of $\it{Conv}(V)\}$;
		\STATE $C^E \gets \{L_i|L_i$ is an edge of $\it{Conv}(V)\}$;
		\STATE $E \gets C^E$;
		\FOR {$P_i$ in $\{P_4,...,P_n\}$}
			\FOR {$P_k$ in $C^P$}
				\IF {$\overline{P_iP_k}$ $\cap$ $C^E$ is $P_k$}
					\STATE $A_i \gets A_i\cup \{P_k\}$;
				\ENDIF
			\ENDFOR
			\STATE $E \gets E\cup\{\overline{P_iP_k}|P_k \in A_i\}$;
			\STATE $V \gets V\cup \{P_i\}$;
			\STATE $C^P \gets \{P_i|P_i$ is the vertex of $\it{Conv}(V)\}$;
			\STATE $C^E \gets \{L_i|L_i$ is the edge of $\it{Conv}(V)\}$;
		\ENDFOR
		\RETURN $E$;
	\end{algorithmic}
\end{algorithm}

According to \cite{8003481}, the time complexity of Algorithm \ref{alg:algorithm1} is $\mathcal{O}(n(\log n)^{\frac{1}{2}})$.

As implied by the definition and construction process, there are no intersecting edges in a non-crossing graph except at the vertices. During the construction process, the number of edges is adaptively adjusted based on the vertex density. In areas with high vertex density, the chances of potential edge intersections increase, thus fewer edges are added to avoid violations. In contrast, in sparse areas with fewer vertices, edge crossing is less likely to occur, so more edges can be incorporated. Evidently, by adapting the edge density according to vertex distribution, a non-crossing graph achieves balanced edge connections for each vertex. The resulting adjacency relations determined by the edges are also balanced naturally. This demonstrates a key advantage of non-crossing graphs -- the ability to construct optimal topology through adaptive edge density control.


We apply non-crossing graphs to construct adjacency relations in SC scenarios based on their characteristics. Specifically, task and worker clusters are modeled as vertices in the non-crossing graph. The edges then determine the adjacencies between clusters - a task cluster is an adjacent task of a worker cluster if directly connected. Symmetrically, the worker cluster has that task cluster as an adjacent task. The adjacent tasks/workers of a given cluster comprise its adjacency list. By extension, k-layer adjacent task/worker clusters are defined as those reachable within k hops.

\begin{figure}[htbp]
	\centering
	\includegraphics[width=0.4\textwidth]{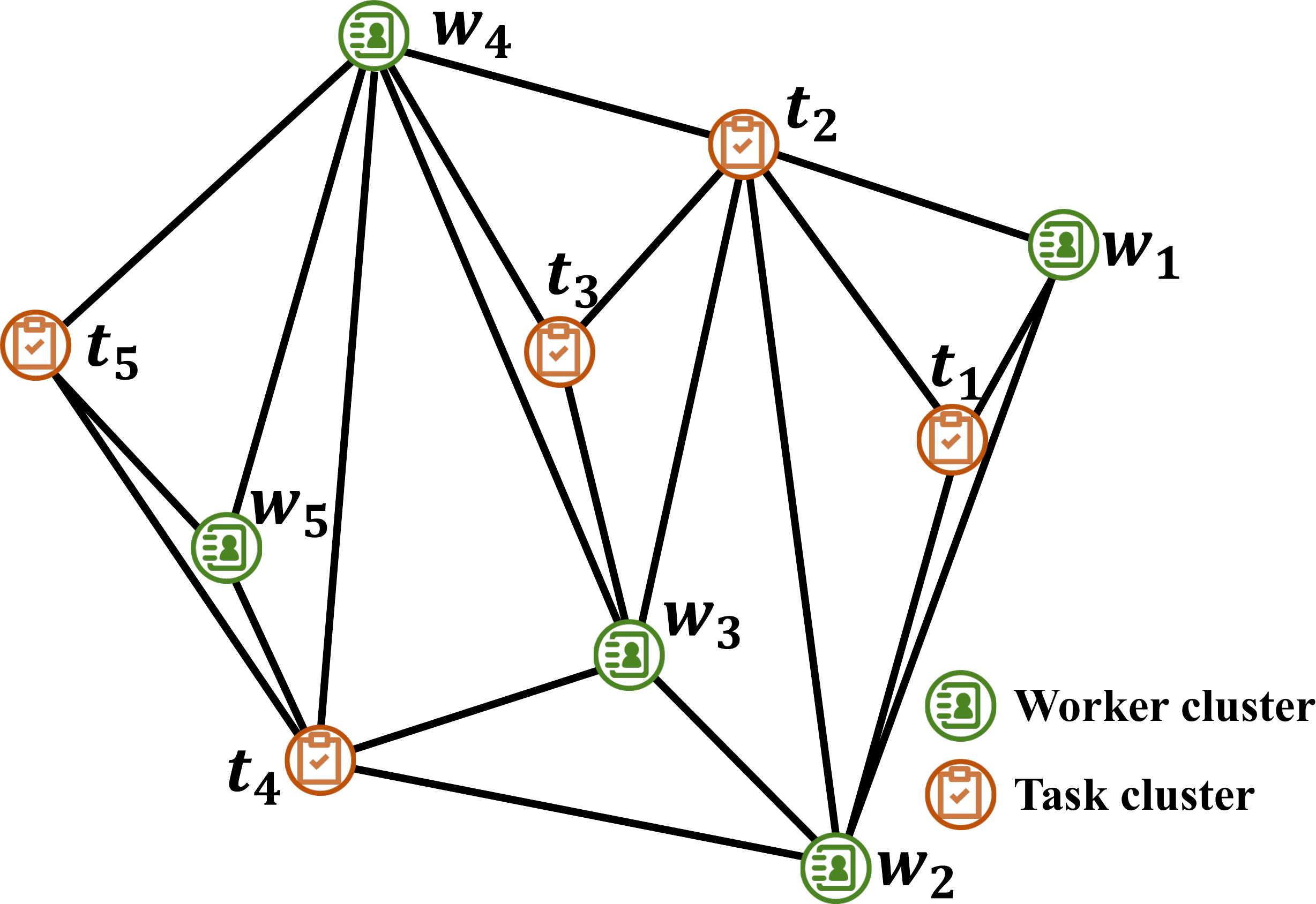}
	\caption{Confused adjacencies}
	\label{fig:confusedadjacency}
\end{figure}

Directly constructing a single non-crossing graph from all worker and task locations can lead to adjacency confusion in spatial crowdsourcing scenarios, as shown in Fig. \ref{fig:confusedadjacency}. For instance, the adjacent nodes of $t_3$ contain both workers and tasks, whereas $w_1$ is not a 1-layer neighbor of $t_1$ despite being a 1-layer adjacent worker. This ambiguity makes it difficult to efficiently query the $L$-layer adjacent clusters for a given worker or task. To address this issue, we propose constructing independent non-crossing graphs per cluster type. Specifically, for each task cluster we build a graph with only worker clusters to obtain its $L$-layer adjacent workers. Symmetrically, independent graphs are generated for every worker cluster to find their $L$-layer adjacent tasks. This approach eliminates adjacency confusion and enables convenient retrieval of layer-wise adjacent nodes.

To insert a new point $P_{new}$ into an existing non-crossing graph $N$, we design Algorithm \ref{alg:algorithm2} to output new edges while minimizing destruction to $N$. The algorithm first divides $N$ into $N_l$ and $N_r$ subgraphs by a vertical line through $P_{new}$, removing intersecting edges (lines 1-4). If $P_{new}$ is external to $N$ (lines 5-10), edges between $P_{new}$ and $N$'s convex hull vertices are added. Otherwise, $P_{new}$ is first inserted into $N_r$ to obtain new edges (lines 12-18). $N_r$'s convex hull and vertex/edge sets are updated (line 20). Finally, $N_l$ and updated $N_r$ are merged into the new graph (lines 21-24).


\begin{algorithm}[t]
	\caption{Insert a new point into the non-crossing graph.}
	\begin{algorithmic}[1]
		\renewcommand{\algorithmicrequire}{\textbf{Input:}}
		\renewcommand{\algorithmicensure}{\textbf{Output:}}
		\label{alg:algorithm2}
		\REQUIRE Non-crossing graphk $N=\{V,E\}$, point to be inserted $P_{new}(x_{new},y_{new})$
		\ENSURE New non-crossing graph $N=\{V,E\}$
		\STATE Draw a vertical line $L_v$ through $P_{new}(x_{new},y_{new})$;
		\STATE $V_{r}\gets \{P_i|P_i\in V$ is to the right of $L_v\}$;
		\STATE $V_{l}\gets \{P_i|P_i\in V$ is to the left of $L_v\}$;
		\STATE $E \gets \{L_j \in E|L_j\cap L_v == \emptyset\}$	
		\IF{ $V_{r}==\emptyset$ \OR $V_{l}==\emptyset$}
			\STATE $C^P \gets \{P_i|P_i$ is the vertex of $\it{Conv}(V)\}$;
			\STATE $C^E \gets \{L_i|L_i$ is the edge of $\it{Conv}(V)\}$;
			\STATE $A \gets \{P_k \in C^P|\overline{P_{new}P_k} \cap C^E$ is $P_k\}$
			\STATE $E \gets E\cup\{\overline{P_{new}P_k}|P_k \in A\}$;
			\STATE $V \gets V\cup \{P_{new}\}$;
		\ELSE
			\STATE $C^P_r \gets \{P_i|P_i$ is the vertex of $\it{Conv}(V_{r})\}$;
			\STATE $C^E_r \gets \{L_i|L_i$ is the edge of $\it{Conv}(V_{r})\}$;
			\STATE $C^P_l \gets \{P_i|P_i$ is the vertex of $\it{Conv}(V_{l})\}$;
			\STATE $C^E_l \gets \{L_i|L_i$ is the edge of $\it{Conv}(V_{l})\}$;
			\STATE $A_{r} \gets \{P_k \in C^P_r|\overline{P_{new}P_k} \cap C^E_r$ is $P_k\}$
			\STATE $E \gets E\cup\{\overline{P_{new}P_k}|P_k \in A_{r}\}$;
			\STATE $V \gets V\cup \{P_{new}\}$;
			\STATE $V_{r} \gets V_{r}\cup \{P_{new}\}$;
			\STATE Update $C^P_r, C^E_r$;
			\FOR {$P_k$ in $C^P_l$}
				\STATE $A_{k} \gets \{P_m \in C^P_r|\overline{P_kP_m} \cap C^E_r = P_m$ \AND $\overline{P_kP_m} \cap C^E_l$ = $P_k\}$;
				\STATE $E \gets E\cup\{\overline{P_kP_m}|P_m \in A_k\}$;
			\ENDFOR
		\ENDIF
		\RETURN $N=\{V,E\}$;
	\end{algorithmic}
\end{algorithm}

We utilize Algorithms \ref{alg:algorithm1} and \ref{alg:algorithm2} to obtain the reconfigured adjacencies for SC as follows:

\begin{enumerate}
	\item Independent non-crossing graphs are constructed for task cluster set $N_t$ and worker cluster set $N_w$ using Algorithm \ref{alg:algorithm1}.
	\item We insert each task cluster $t_i$ from $N_t$ into the worker graph $N_w$ via Algorithm \ref{alg:algorithm2}, which outputs the $L$-layer adjacent worker clusters of $t_i$.
	\item Symmetrically, every worker cluster $w_j$ from $N_w$ is inserted into the task graph $N_t$ to obtain the $L$-layer adjacent tasks of $w_j$.
\end{enumerate}

\begin{figure}[htbp]
	\centering
	\includegraphics[width=0.45\textwidth]{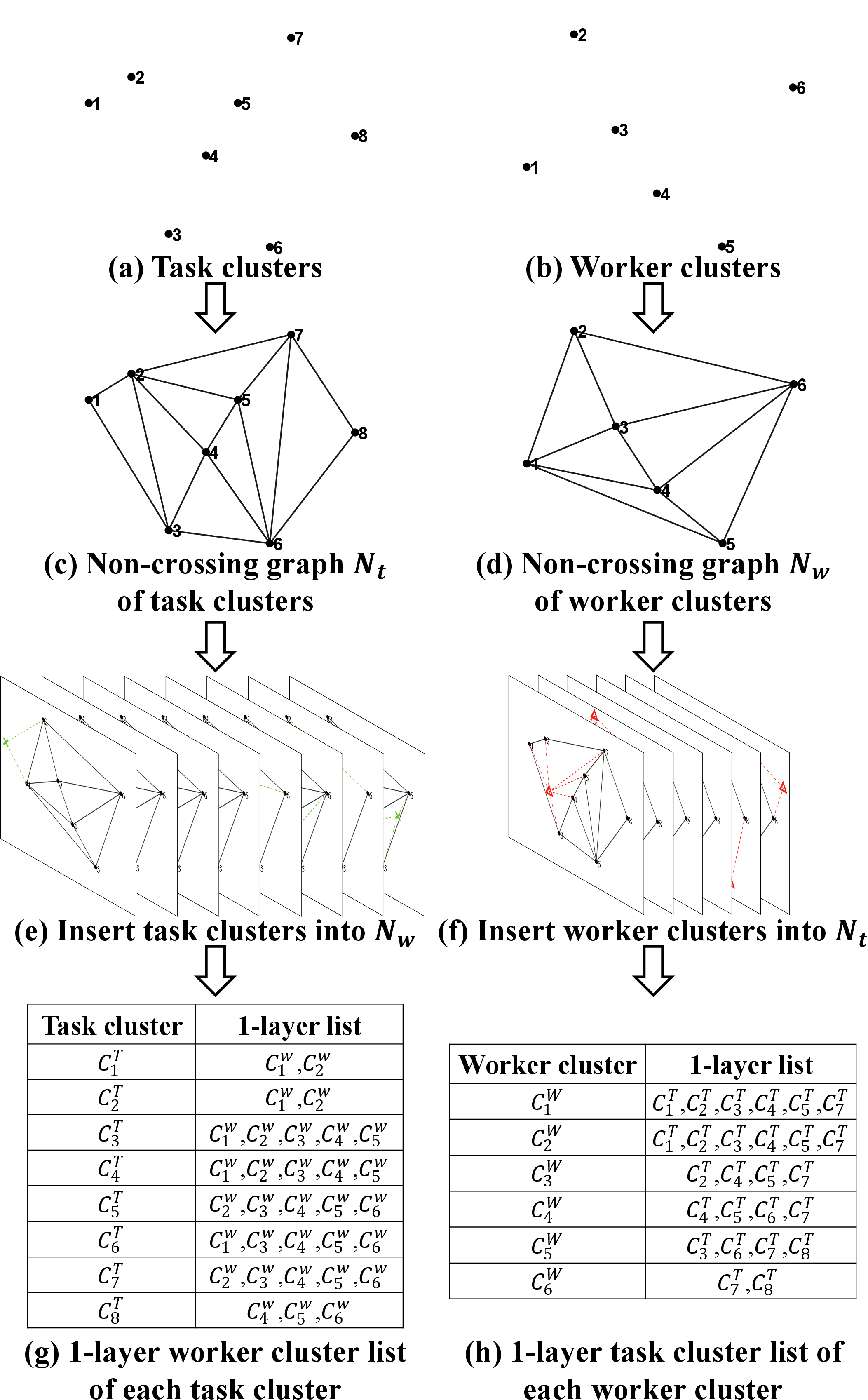}
	\caption{Adjacencies reconfiguration process.}
	\label{fig:adjprocess}
\end{figure}

By repeating Steps 2 and 3 for all task and worker clusters, the reconfigured adjacency lists are generated. These updated adjacencies can then be utilized for optimized task allocation. As an example, Fig. \ref{fig:adjprocess} illustrates the 1-layer reconfiguration process on a scenario with 8 task and 6 worker clusters randomly located (as shown in Fig. \ref{fig:adjprocess}(a) and (b)). Specifically, the reconfiguration involves:
\begin{enumerate}
	\item Constructing independent task and worker graphs $N_t$ and $N_w$ using Algorithm \ref{alg:algorithm1}(Fig. \ref{fig:adjprocess}(c) and (d)).
	\item Inserting each task cluster into $N_w$ via Algorithm \ref{alg:algorithm2} (Fig. \ref{fig:adjprocess}(e)).
	\item Symmetrically inserting every worker cluster into $N_t$ (Fig. \ref{fig:adjprocess}(f)).
	\item Acquiring the 1-layer adjacent clusters for each task and worker cluster through the edges in the corresponding non-crossing graphs.
\end{enumerate}

\section{Allocation module}\label{sec:allocation}
The allocation module designed in this paper consists of the evaluation module and the bi-directional matching module, which implement the evaluation of a task cluster and a worker cluster and the bi-directional matching based on the evaluation results, respectively, and finally obtain the task allocation scheme.

\subsection{Evaluation module}
We evaluate cluster quality based on two metrics: value and dispersion. The value represents the utility a cluster provides for its constituent tasks or workers. Dispersion indicates the deviation of elements within a cluster from its center. Higher value and lower dispersion are desirable cluster traits. Additionally, as workers must travel to task locations, inter-cluster distance between task and worker clusters is also pertinent for evaluation.

We assess each task cluster $C^T_i$ for a given worker cluster $C^W_{any}$ using the evaluation function $E_t()$ defined in Eq. (\ref{eq:et}). This contains three components:
\begin{itemize}
	\item $p(C^T_i)$ representing the utility of $C^T_i$ for $C^W_{any}$, calculated based on individual task benefits in $C^T_i$.
	\item $D(C^T_i)$ measuring the dispersion of $C^T_i$, determined by the maximum distance between any two constituent tasks.
	\item The distance between centroids of $C^T_i$ and $C^W_{any}$. Shorter distances imply faster travel and less cost for workers.
\end{itemize}
The coefficients $\alpha$, $\beta$, and $\gamma$ weight the three components based on context. Higher values of $p()$ and lower values of $D()$ and distance improve the score $E_t()$.
\begin{equation}
E_t(C^T_i)=\alpha p(C^T_i)-\beta D(C^T_i)-\gamma d(C^T_i,C^W_{any}) \label{eq:et}
\end{equation}


Analogously, we assess each worker cluster $C^W_i$ against a given task cluster $C^T_{any}$ using the function $E_w()$ defined in Eq. (\ref{eq:ew}). This comprises:
\begin{itemize}
	\item $a(C^W_i)$ denoting the utility of $C^W_i$ for $C^T_{any}$, positively correlated with workers' abilities. For simplicity, we equate the two in subsequent analysis.
	\item $D(C^W_i)$ measuring the dispersion of $C^W_i$, based on the variance of constituent workers' abilities.
	\item The distance between $C^W_i$ and $C^T_{any}$, as in Eq. (\ref{eq:et}).
\end{itemize}
The coefficients weight these components depending on the context. Higher values of $a()$ and lower values of $D()$ and distance improve the overall score $E_w()$.
\begin{equation}
E_w(C^W_i)=\alpha a(C^W_i)-\beta D(C^W_i)-\gamma d(C^W_i,C^T_{any}) \label{eq:ew}
\end{equation}

\begin{figure}[htbp]
	\centering
	\includegraphics[width=0.3\textwidth]{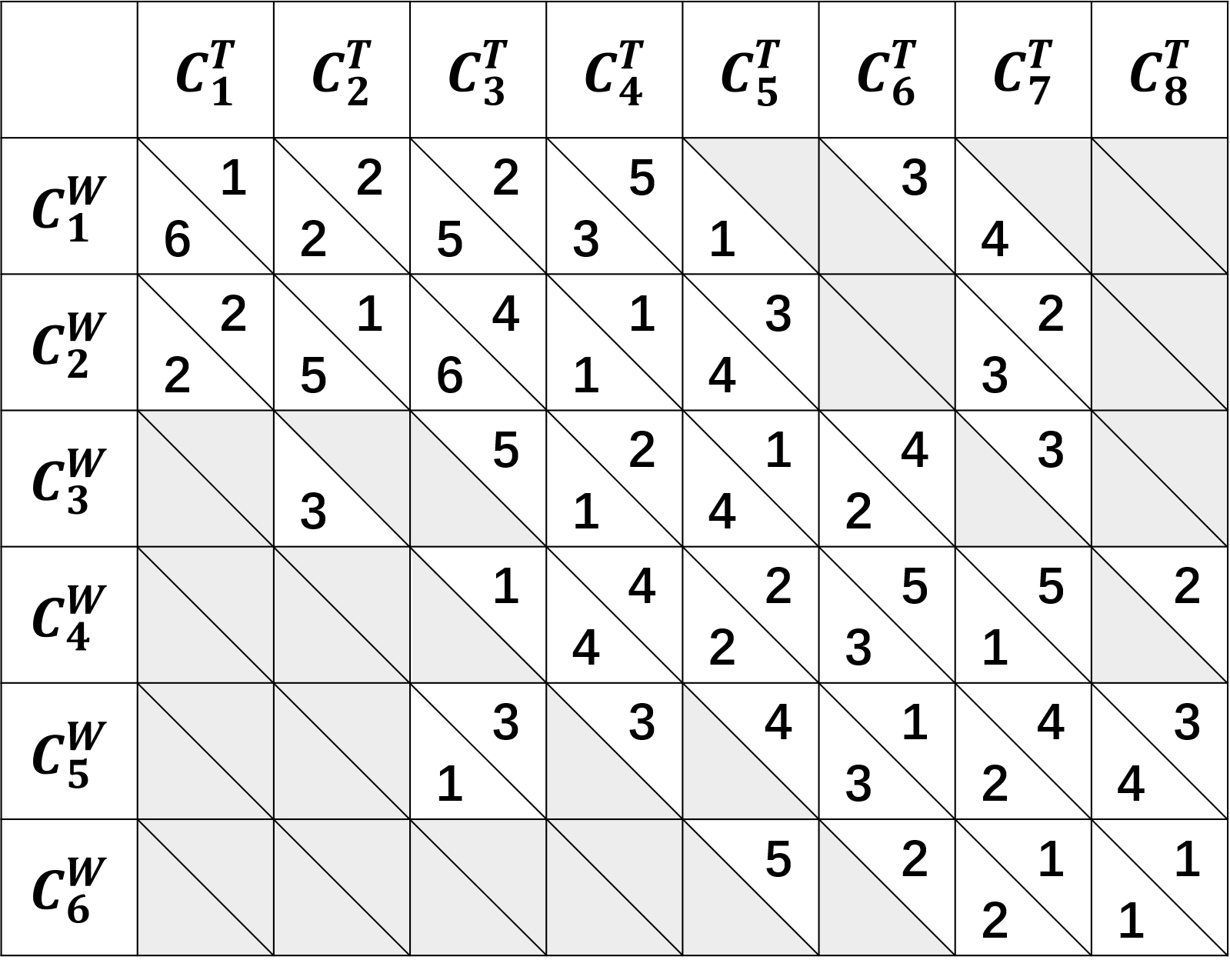}
	\caption{Table of adjacencies with evaluation.}
	\label{fig:adjlist}
\end{figure}

We rank the $L$-layer adjacent worker/task clusters for each task/worker cluster by sorting their evaluation scores $E_t()$ and $E_w()$  in descending order. This generates adjacency lists with ranked entries. As an example, Fig. \ref{fig:adjlist} tabulates the evaluation-based rankings for the sample adjacencies in Fig. \ref{fig:adjprocess}. The number in row $i$ column $j$ denotes the rank of task cluster $C^T_j$ in worker cluster $C^W_i$'s adjacency list. Symmetrically, the number in row $j$ column $i$ indicates the rank of $C^W_i$ in $C^T_j$'s list.

Notably, using total or average values of a cluster's components in $E_t()$ and $E_w()$ reflects different objectives, leading to varying allocation schemes. We experimentally analyze this distinction in cluster value orientations later.

\subsection{Matching module}
\begin{figure}[htbp]
	\centering
	\includegraphics[width=0.45\textwidth]{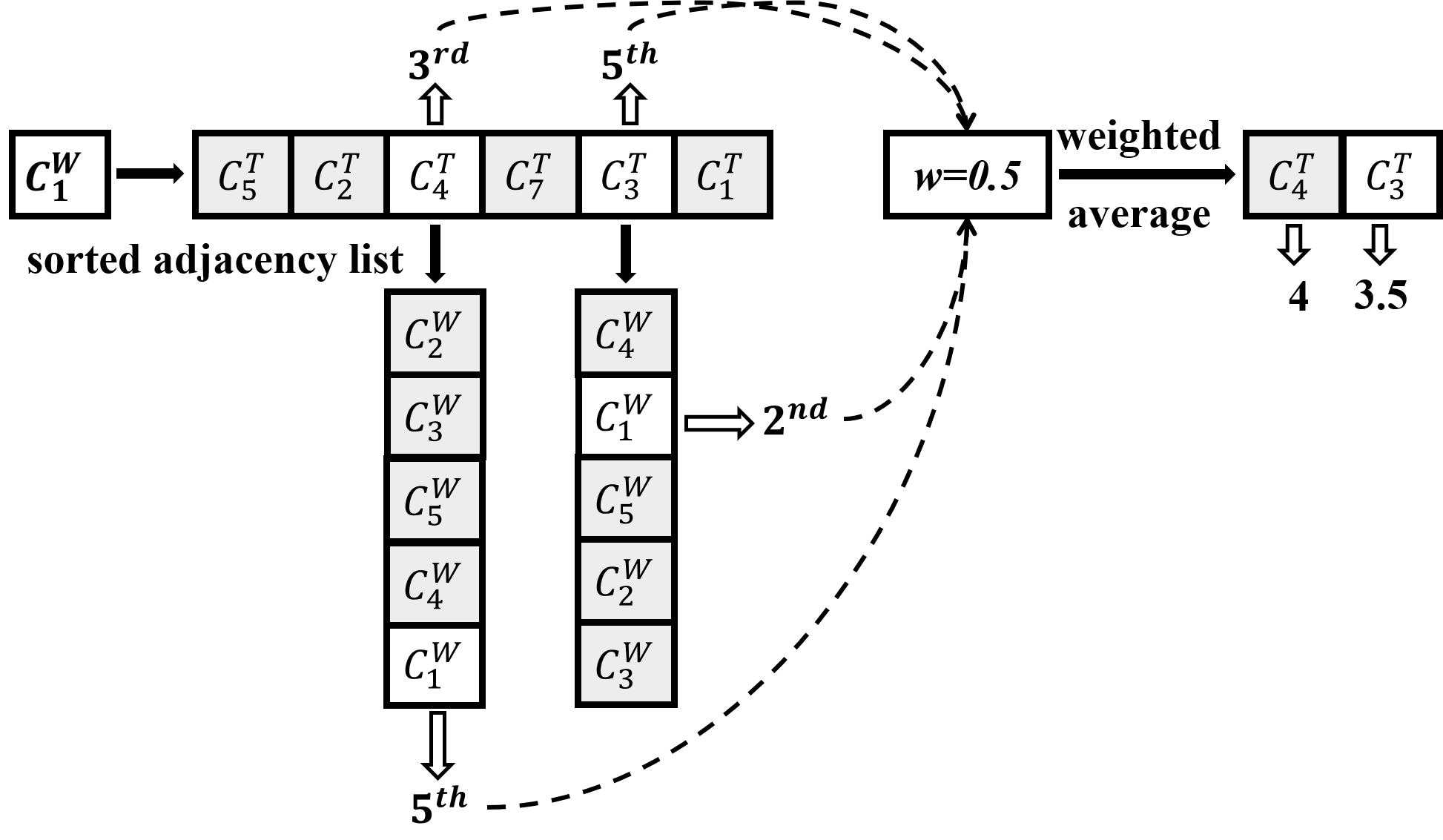}
	\caption{Diagram of the adjacency merging process.}
	\label{fig:merge}
\end{figure}

\begin{figure}[htbp]
	\centering
	\includegraphics[width=0.3\textwidth]{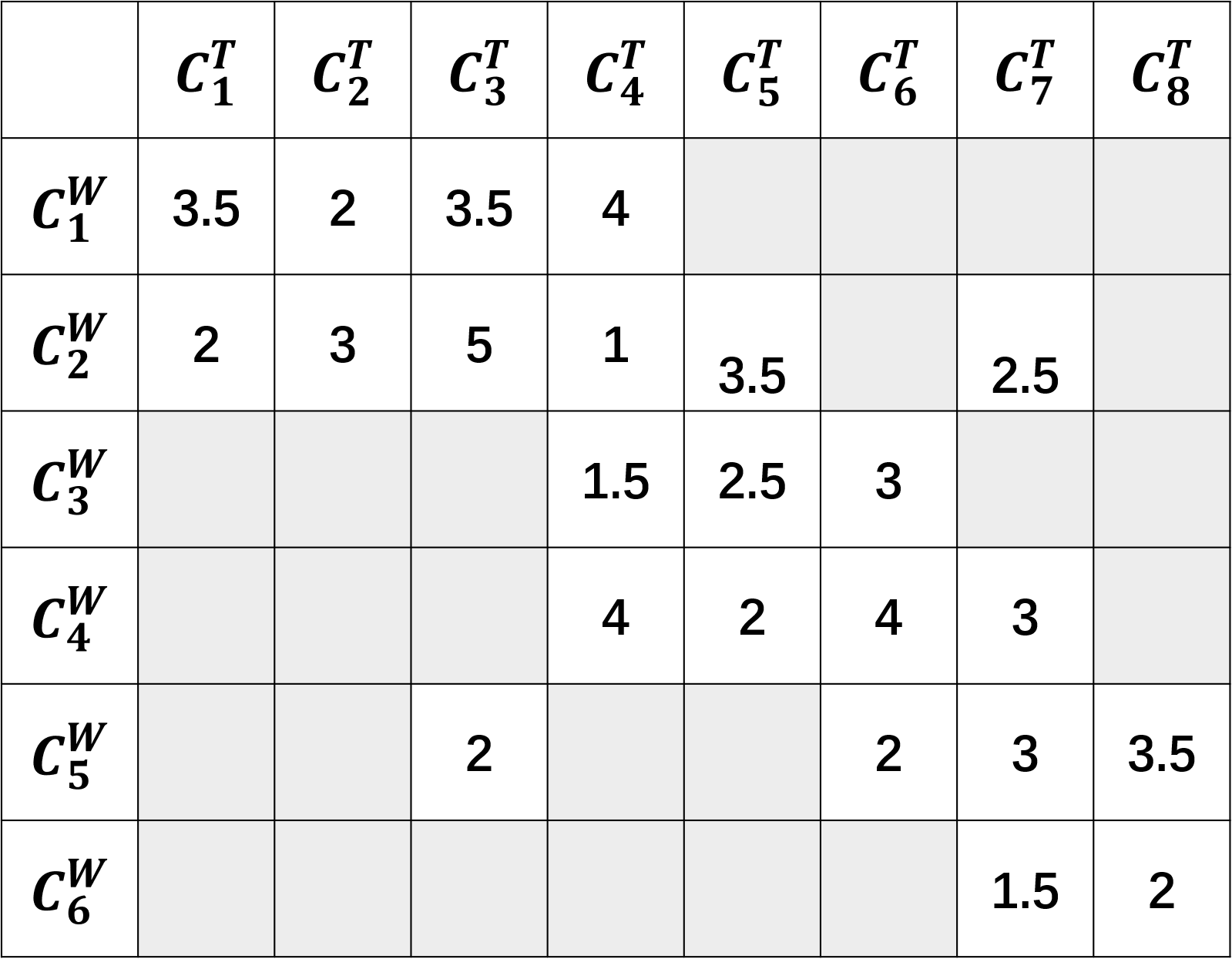}
	\caption{Diagram of the matching table.}
	\label{fig:matchingtable}
\end{figure}

In the matching module, we merge the ranked adjacency lists using Eq. (\ref{eq:merge}). Here $rank_t$ and $rank_w$ denote the ranks of the currently paired task and worker clusters in each other's lists. The weights $w$ and $(1-w)$ balance their contributions. The result is the merged rank $rank$. Fig. \ref{fig:merge} illustrates this for worker cluster $C^W_1$ from Fig. \ref{fig:adjprocess}. The merged adjacency lists form a matching table, with entries called matching values. Lower matching values indicate better task-worker match quality. Fig. \ref{fig:matchingtable} shows the matching table for our example, derived by merging the ranked adjacency lists.

Task matching is performed from a worker perspective to better account for workers' individual movement costs. Specifically, worker clusters are traversed in a certain order. For each worker cluster $C^W$, the task cluster with the minimum matching value is batch-allocated to the workers in $C^W$. This worker-centric scheme allows preferentially matching proximate tasks to workers based on their positions.

\begin{equation}
rank = w*rank_t+(1-w)*rank_w \label{eq:merge}
\end{equation}
The above allocation scheme does not limit the number of assignments per task. However, real-world crowdsourcing often constrains total allocation rounds. Moreover, unlimited assignments can result in top-ranked tasks being repeatedly allocated, making it difficult to match poorer tasks. Hence, during bi-directional matching, we exclude task clusters already assigned to multiple workers from subsequent rounds. If a worker cluster's adjacent task clusters all reach the allocation limit, the worker cluster will not be matched further in the current round.

Under this allocation scheme, the traversal order of worker clusters clearly impacts the resulting task allocation. We experimentally analyze the effects of various traversal orders on allocation outcomes later.

\section{Performance evaluation}\label{sec:evaluation}
In this section, we utilize a mix of real-world and simulated data to evaluate our proposed task allocation scheme under varying parameter settings. Real-world POI and driver trajectory data from ride-hailing applications provide the spatial distribution of tasks and workers. Corresponding simulation data such as task rewards and worker abilities are generated based on normal distributions. After preprocessing, we conduct extensive experiments to analyze the performance of our allocation scheme using seven key indicators detailed below.

\subsection{Data preprocessing and parameter settings}
In spatial crowdsourcing scenarios, due to the dynamic nature of workers’ locations, we can only obtain their trajectory data over time instead of direct position data. According to human mobility patterns\cite{barabasi2005origin}, human spatial motion exhibits significant locality - the probability of being far from a small area decays slowly over time. For individuals, there are usually a few frequently visited locations. In other words, most people's trajectory of motion is within a certain area. Therefore, we propose clustering the trajectory data of each worker and using the cluster center as the representative location of that worker. This allows us to approximate worker geography from trajectory information despite mobility.

The worker trajectory data is derived from anonymized online ride records from Didi's GAIA initiative, comprising one month of order and mobility data in northeast Chengdu, China (lat 30.65 to 30.72°N, long 104.04 to 104.12°E). As the raw data includes anomalies such as canceled orders and duplicates from collection errors, preprocessing involves deduplication and cleansing. We cluster trajectories per driver ID and identify each driver's location using the central coordinate of the minimum enclosing circle of their trajectory cluster.

The task location data is derived from 96,947 anonymized POIs from Didi's GAIA initiative encompassing Chengdu, China (30.59°N to 30.73°N, 103.99°E to 104.17°E). We extract the geographic coordinates corresponding to the region covered by the worker trajectory data (30.65°N to 30.72°N, 104.04°E to 104.12°E) to align task and worker locations.

Since the datasets lack actual order prices, ratings, and other crowdsourcing attributes, we simulate such data by randomly generating task rewards and worker abilities following normal distributions. This yields the necessary inputs for calculating optimal allocations based on our proposed approach.

With limited task allocations, varying the order of worker cluster traversal impacts the allocation outcomes. We design five traversal schemes:
\begin{itemize}
	\item NonLMT: Unlimited allocations, so traversal order does not affect results.
	\item Random: Worker clusters are traversed in a random order.
	\item Xcoord: Worker clusters are traversed in descending order of $x$-coordinates.
	\item AVG: Worker clusters are traversed in descending order of the average ability of workers in the cluster.
	\item SUM: Worker clusters are traversed in descending order of the sum ability of workers in the cluster.
\end{itemize}

Using total or average values when evaluating task and worker clusters also impacts allocations. We design four evaluation schemes:
\begin{itemize}
	\item AVG-AVG: For both task and worker clusters, use average value of components. i.e., $p(C^T)$ and $a(C^W)$ are average task benefits and worker abilities.
	\item SUM-SUM:For both task and worker clusters, use sum of component values. i.e., $p(C^T)$ and $a(C^W)$ are sums of task benefits and abilities.
	\item AVG-SUM: For task clusters, use average tasks' benefits. For worker clusters, use sum of workers' abilities.
	\item SUM-AVG: For task clusters, use sum of tasks' benefits. For worker clusters, use average of workers' abilities.
\end{itemize}

Additionally, parameters like the number of adjacent layers and weight $w$ in bi-directional matching impact the allocation outcomes. We conduct a two-stage experiment to comprehensively analyze performance under various settings: in stage 1, we set $w=0.5$ and evaluate schemes for all combinations of different traversal modes, evaluation bases, and 1-layer versus 2-layer adjacent node selections. In stage 2, we test the effect of varying $w$ on the optimal scheme from stage 1 in terms of adjacent layers, traversal mode, and evaluation basis. In both stages, the task allocation limit is 15 rounds for all traversal modes except NonLMT.

\subsection{Evaluation results}
To analyze performance from multiple angles, we evaluate the task allocation schemes along six dimensions:
\begin{itemize}
	\item Task allocation rate: the ratio of allocated tasks to total tasks.
	\item Worker utilization rate: the ratio of workers assigned tasks to total workers. 
	\item Total requester payoff: the sum of rewards gained by all requesters.
	\item Total worker payoff: the total payments received by workers.
	\item Requester payoff variance: reward disparity among requesters.
	\item Worker payoff variance: payment disparity among workers.
\end{itemize}
The first two metrics assess allocation efficiency. The third and fourth reflect impacts on social welfare. The last two indicate fairness and stability. Together, these provide a comprehensive evaluation to select the optimal scheme.

\begin{figure}[htbp]
	\centering
	\includegraphics[width=0.48\textwidth]{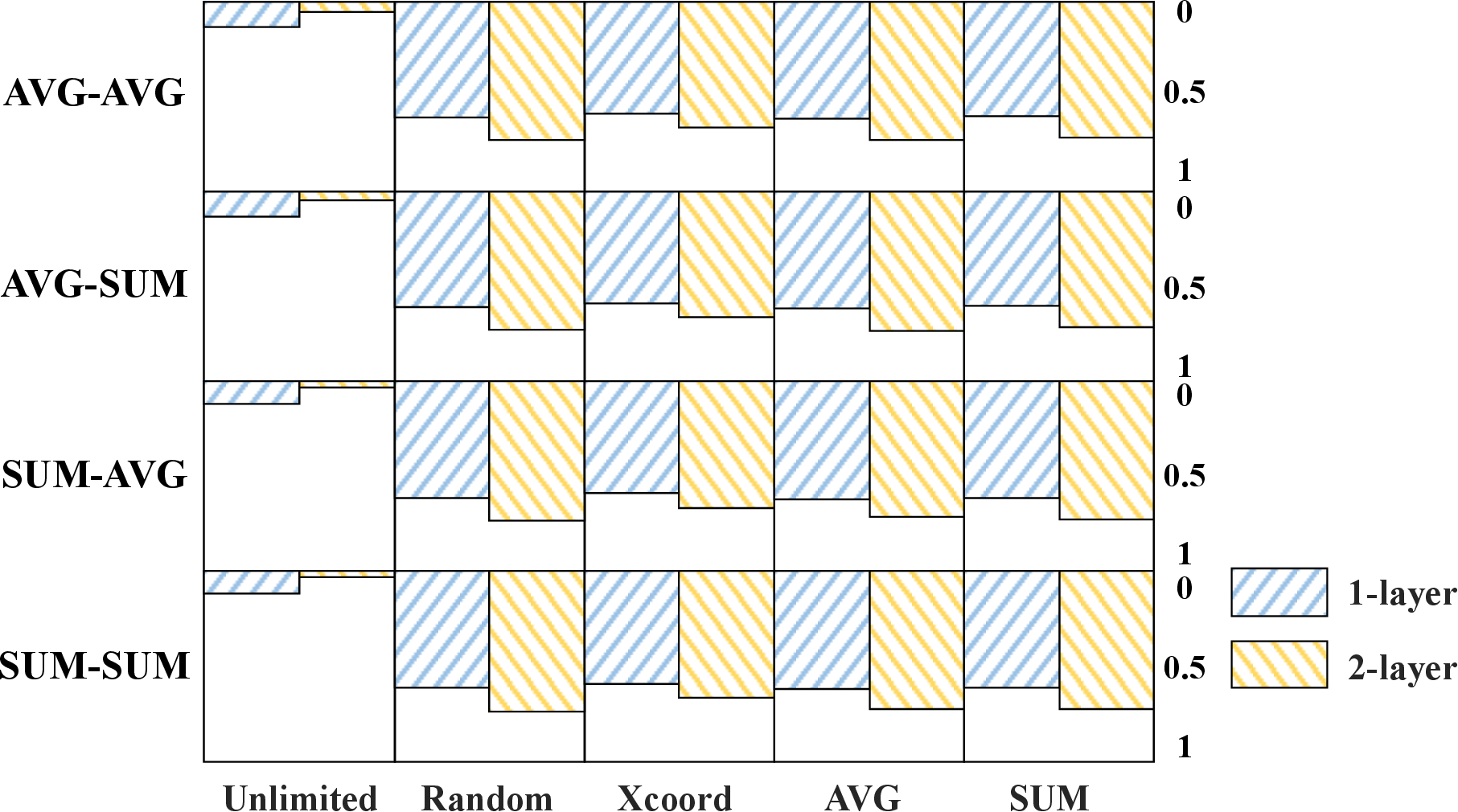}
	\caption{Task allocation rate diagram.}
	\label{fig:taskratio}
\end{figure}

\begin{figure}[htbp]
	\centering
	\includegraphics[width=0.48\textwidth]{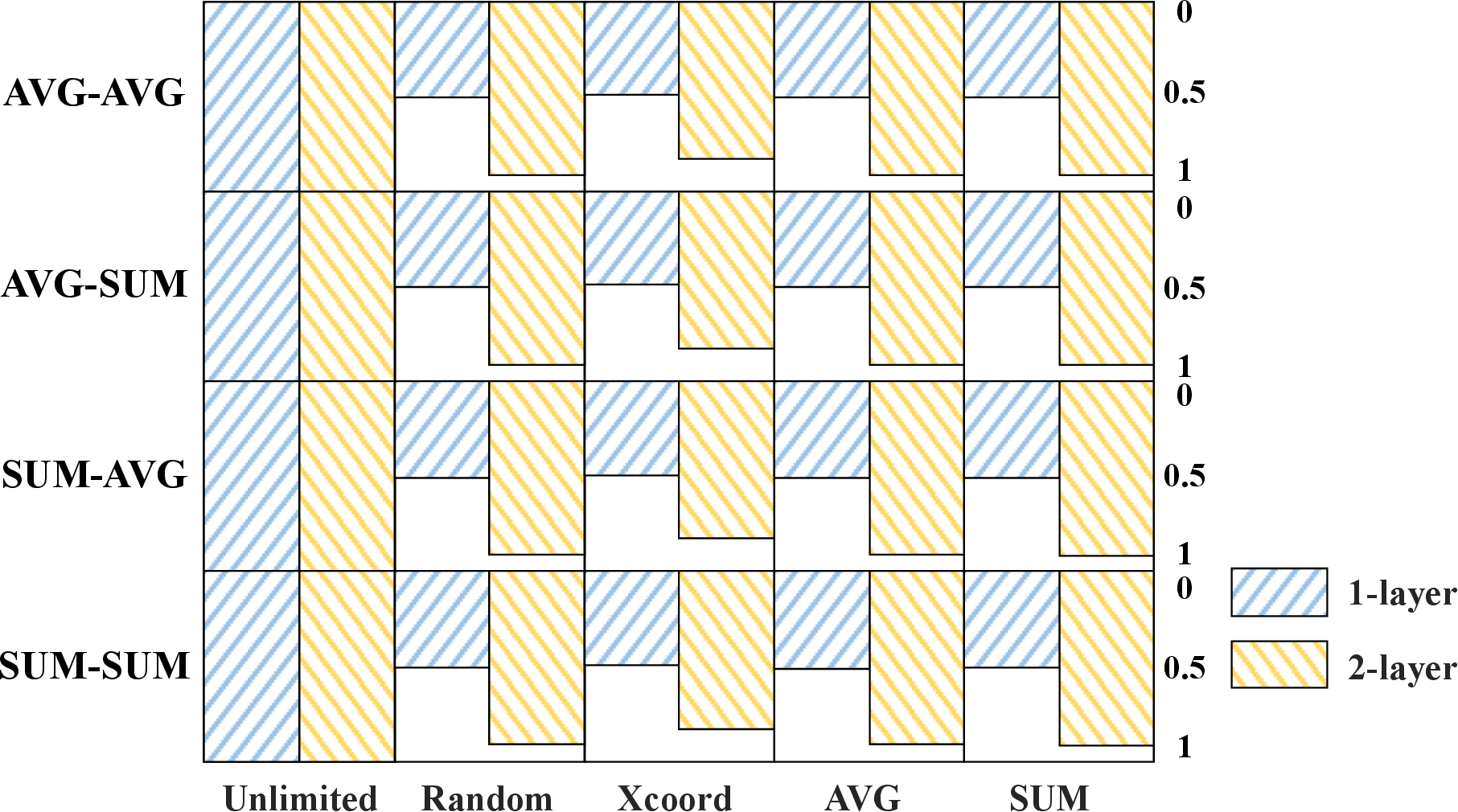}
	\caption{Workers allocation rate diagram.}
	\label{fig:workerratio}
\end{figure}


Figure \ref{fig:taskratio} illustrates the task allocation rates for schemes with varying adjacency layers, worker traversal orders, and evaluation bases. The color and texture of the rectangular blocks denote the number of adjacency layers. The block height as a percentage of the grid height indicates the allocation rate. In the absence of limits on the number of task allocations, the rates are under 20\% for all schemes. After imposing a limit of 15 allocation rounds, the rates improve to 50\% or higher, demonstrating the efficacy of this constraint for increasing task allocation.


Comparing allocation rates shows 2-layer adjacency improves performance given limited allocations per task, while 1-layer is better without limits. With more layers, each task appears in more worker adjacency lists, increasing its chances of allocation. However, tasks also face more competitors. Unrestricted allocations often cause low-scoring tasks to fail allocation, reducing the allocation rate. In contrast, limits on allocations exclude top tasks after sufficient allocations, removing them as competitors for remaining tasks.


Limiting per-task allocations implies some worker clusters may remain unallocated. We compare worker utilization rates for schemes in Figure \ref{fig:workerratio}. The 2-layer schemes markedly outperform 1-layer designs. This results from the expanded adjacency lists under 2-layer graphs, providing more worker-task pairing opportunities.

The similar allocation and utilization rates across evaluation bases in Figures \ref{fig:taskratio} and \ref{fig:workerratio} indicate minimal impact from this parameter. However, traversal mode effects are visible - Xcoord performs relatively poorly on both metrics. This scheme's reliance on spatial distributions causes failure when worker clusters with fewer tasks are traversed later after their candidate tasks hit the limit. For the data distribution in this experiment, Xcoord encounters such scenarios, exemplifying the risks of strong spatial priors.

\begin{figure}[htb]
	\centering  
	\subfigure[Workers' payoff.]{
		\label{fig:indicator1}
		\includegraphics[width=0.45\linewidth]{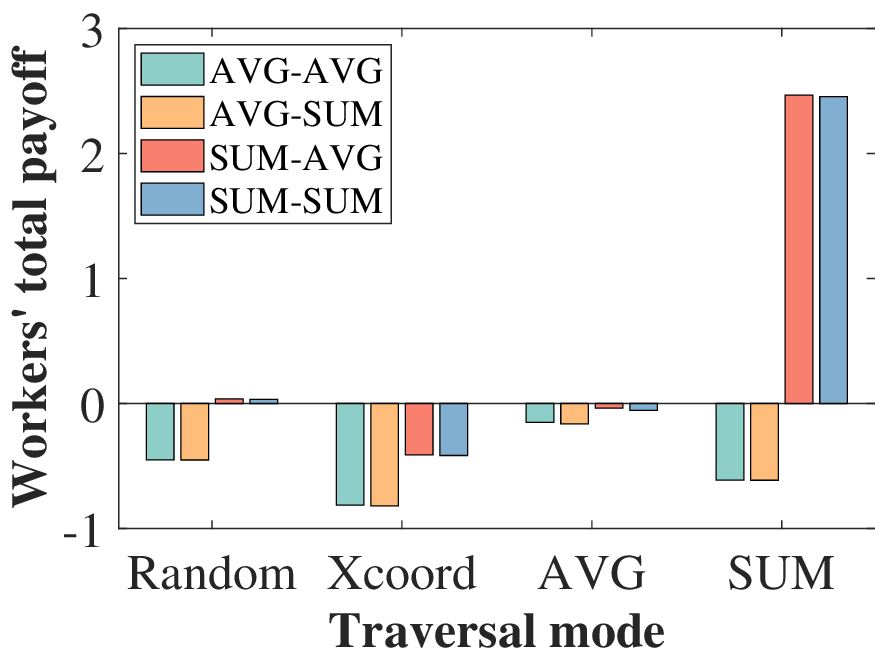}}		
	\subfigure[requesters' payoff.]{
		\label{fig:indicator2}
		\includegraphics[width=0.45\linewidth]{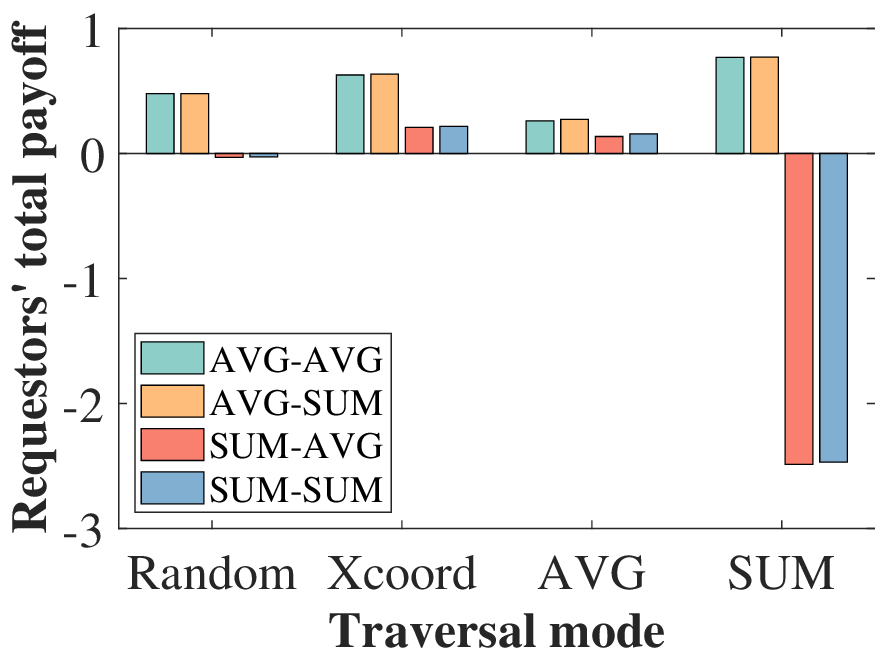}}		
	  \\
	\subfigure[Variance of workers' payoff.]{
		\label{fig:indicator3}
		\includegraphics[width=0.45\linewidth]{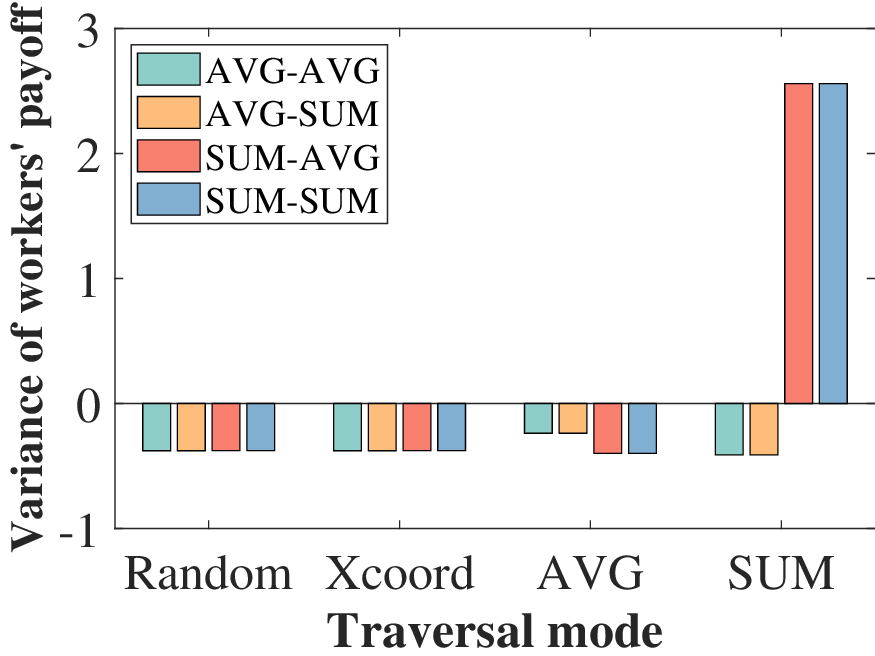}}		
	\subfigure[Variance of requesters' payoff.]{
		\label{fig:indicator4}
		\includegraphics[width=0.45\linewidth]{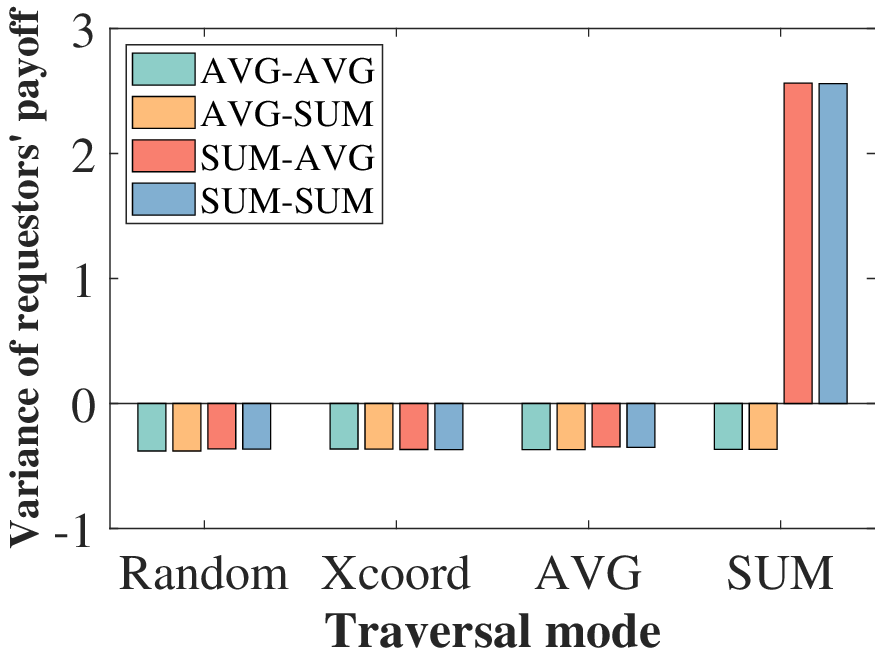}}		
	\caption{Schematic of payoff and payoff variance.}
	\label{fig:indicators}
\end{figure}

For 2-layer schemes, we further analyze requester and worker payoffs and variances. As this work focuses on cross-scheme comparisons rather than absolute values, we standardize the four metrics in Figure \ref{fig:indicators} to minimize magnitude differences and attribute effects. Random-order traversal is used as the baseline. Standardization enables normalized comparisons, with random traversal as a consistent reference across indicator categories.

Figures \ref{fig:indicator1} and \ref{fig:indicator2} show Xcoord traversal reduces total worker payoff while increasing requester payoff versus random baseline. AVG and SUM traversal depend on the evaluation basis. 

The SUM-AVG and SUM-SUM evaluation bases in Figure \ref{fig:indicator1} maximize total worker payoff, reflecting a greedy approach - higher valued tasks are prioritized for each worker cluster. Comparing traversals shows ordering by total cluster ability improves worker payoff. With normally distributed abilities, higher totals often indicate more workers, so allocating top tasks to larger clusters amplifies payoff versus other schemes.
Analyzing requester payoff in Figure \ref{fig:indicator2} reveals AVG-SUM and SUM-SUM evaluation bases yield higher returns regardless of traversal mode. This aligns with greedy task prioritization favoring higher total value clusters.

Figures \ref{fig:indicator3} and \ref{fig:indicator4} illustrate variance in worker and requester payoffs under different schemes. Overall, traversal mode has minimal impact, indicating variance depends on other factors like evaluation basis. SUM-AVG and SUM-SUM evaluation focuses on total cluster value, allocating high-revenue tasks to large, skilled worker clusters. This concentration widens payoff disparities for both groups.

\begin{figure}[htb]
	\centering  
	\subfigure[Schematic of the effect of weights on the allocation rates.]{
		\label{fig:weight1}
		\includegraphics[width=0.45\linewidth]{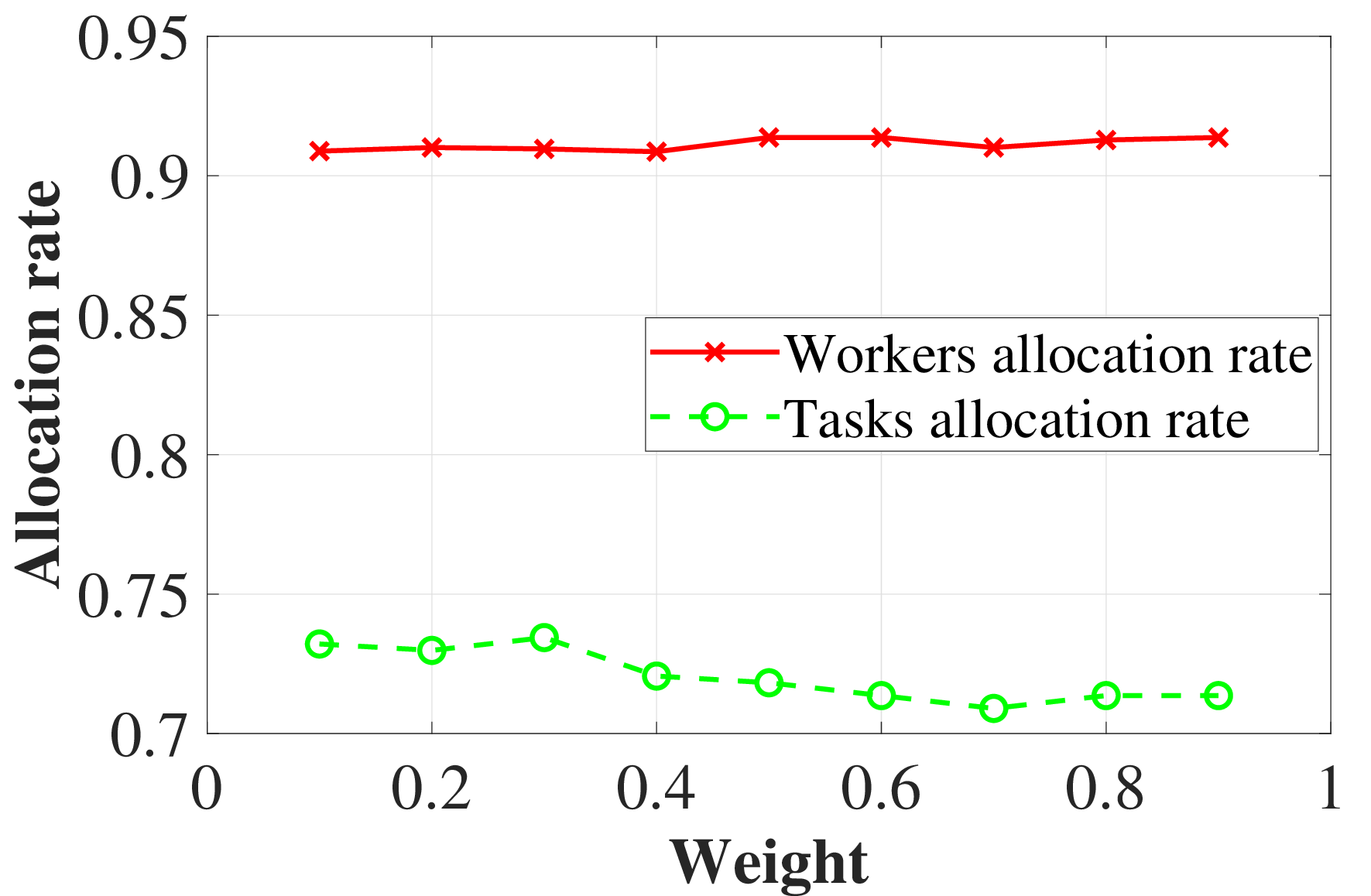}}
	\subfigure[Schematic of the effect of weights on the payoffs.]{
		\label{fig:weight2}
		\includegraphics[width=0.45\linewidth]{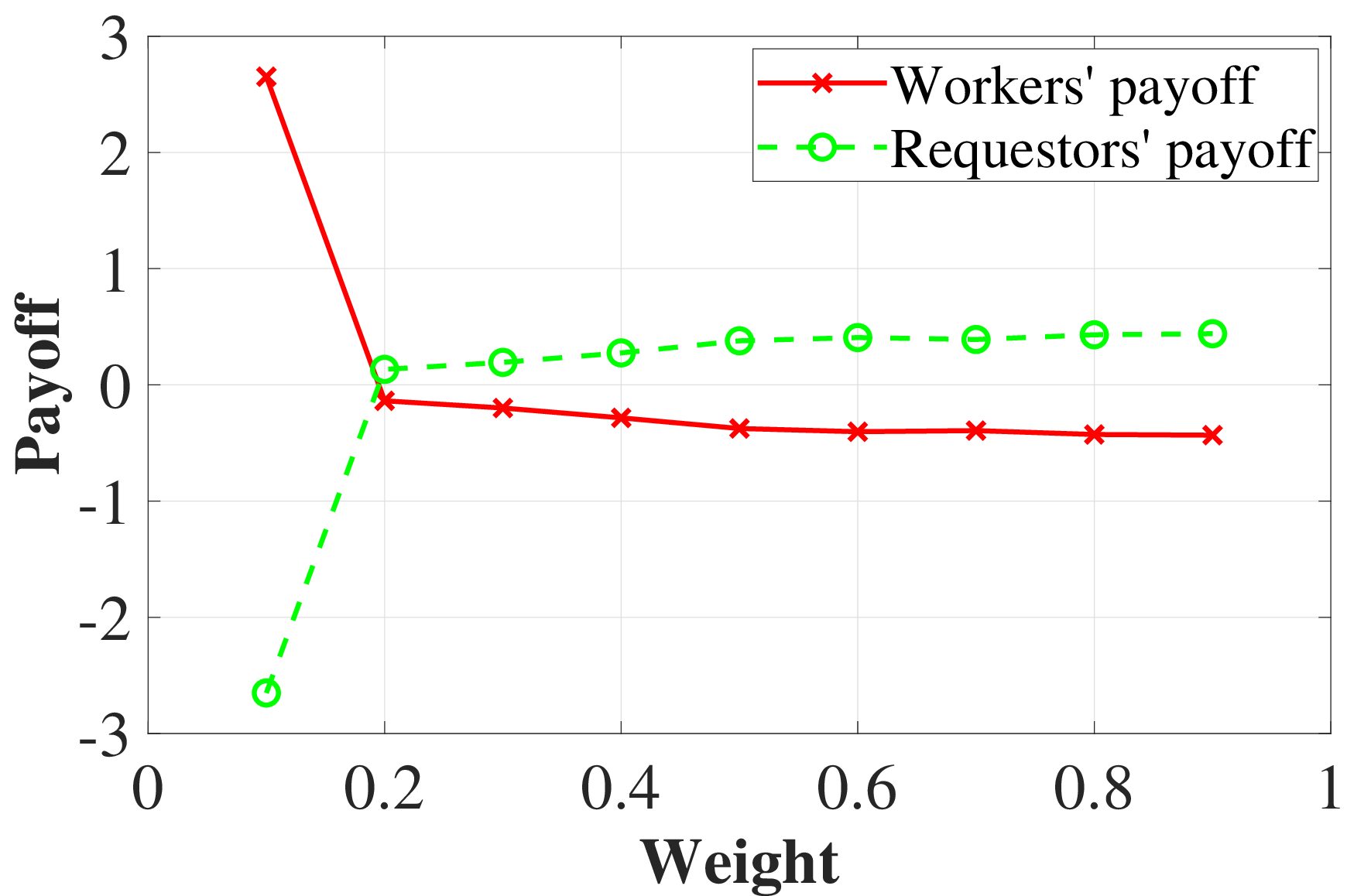}}		
	\caption{The influence of different $w$ on the allocation scheme.}
	\label{fig:weights}
\end{figure}

We further explore the matching weight $w$ in 2-layer AVG-AVG traversal and SUM evaluation schemes (Figure \ref{fig:weights}). The impact on allocation and matching rates shows adjacency layers dominate. As $w$ increases, requester payoff rises while worker payoff declines, since higher $w$ weights task-side evaluations more, reducing worker earnings.

\section{Conclusion}\label{sec:conclusion}
In this work, we propose a spatial crowdsourcing task allocation framework to handle massive spatial data, where clustering helps reduce computational complexity for large-scale inputs. We introduce non-crossing graphs and associated reconfiguration algorithms for optimized matching. Our bi-directional assignment approach jointly considers requester and worker benefits. Experiments on real-world data demonstrate increased task allocation rate and worker utilization rate. Experimental results align with intuitive reasoning: 1) Limited allocations cause adjacency layers to dominate rates, minimizing other factors; 2) Prioritizing total task cluster value improves worker payoffs, while total worker cluster ability raises requester earnings; 3) Focusing on total task value also increases payoff variances; 4) Higher weights for task-side evaluations in bi-directional matching increase requester payoffs at the expense of worker earnings. This paper focuses solely on geographic data for simplicity. Future work will expand the matching schemes to incorporate complex constraints like temporal constraints and task attributes.

\bibliographystyle{IEEEtran}
\bibliography{reference}
\begin{IEEEbiography}[{\includegraphics[width=1in,height=1.25in,clip,keepaspectratio]{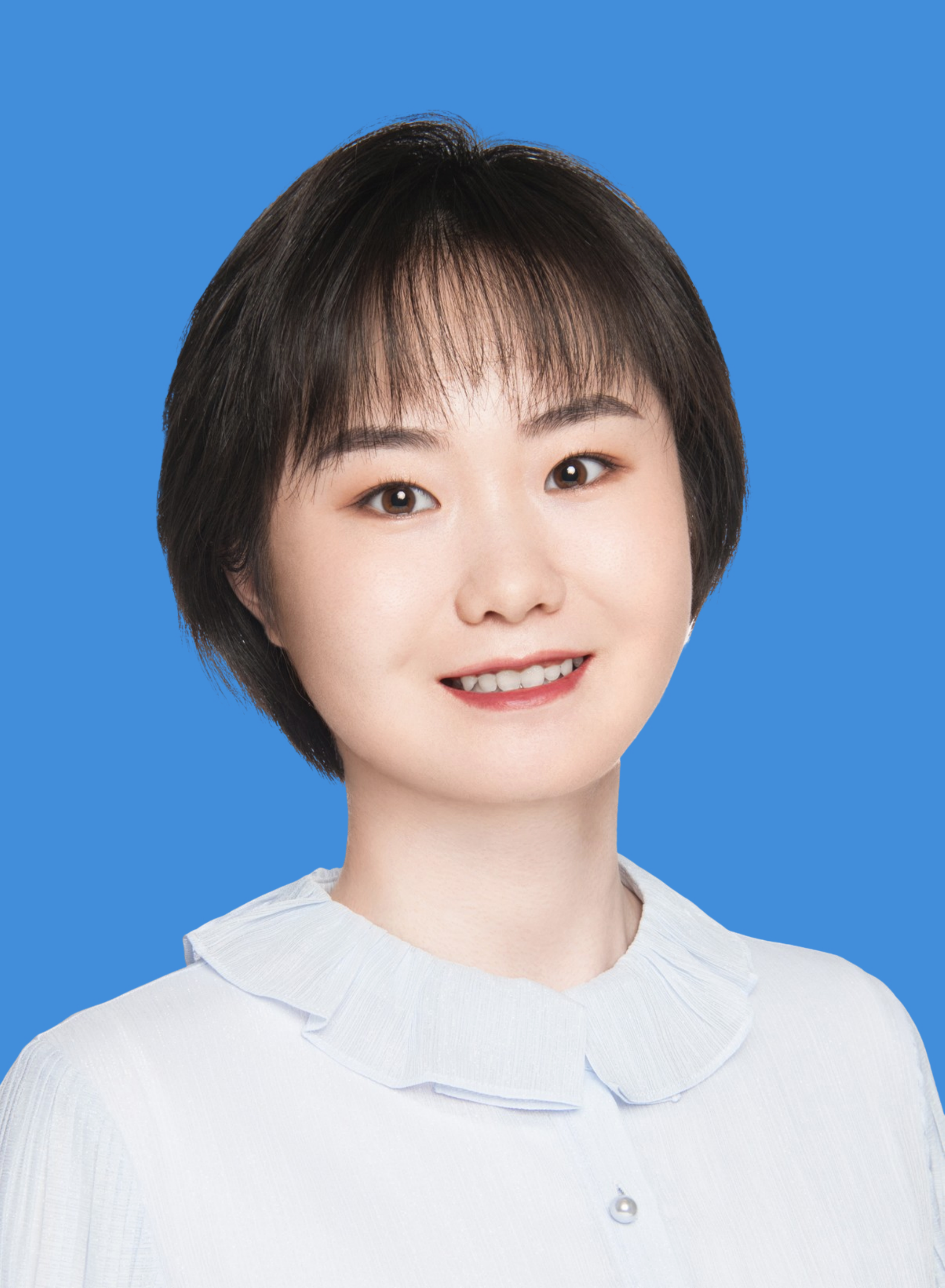}}]{Kun Li}
	received the Ph. D. degree in 2023 from  School of Artificial Intelligence, Beijing Normal University. Now, she is an Assistant Professor at Shandong University. Her research interests include crowdsourcing, mobile computing, and blockchain.
\end{IEEEbiography}
\begin{IEEEbiography}[{\includegraphics[width=1in,height=1.25in,clip,keepaspectratio]{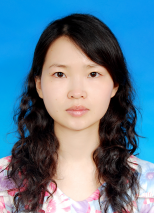}}]{Shengling Wang}
	is a full professor in the School of Artificial Intelligence, Beijing Normal University. She received her Ph.D. in 2008 from Xi'an Jiaotong University. After that, she did her postdoctoral research in the Department of Computer Science and Technology, Tsinghua University. Then she worked as an assistant and associate professor from 2010 to 2013 in the Institute of Computing Technology of the Chinese Academy of Sciences. Her research interests include mobile/wireless networks, game theory, crowdsourcing.
\end{IEEEbiography}
\begin{IEEEbiography}[{\includegraphics[width=1in,height=1.25in,clip,keepaspectratio]{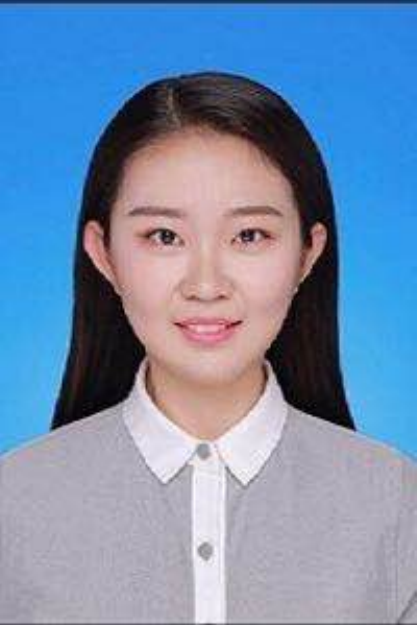}}]
	{Hongwei Shi}
	reccived her Ph.D. degree in Computer Science from Beijing Normal University in 2023.Now she is a lecturer in Beijing Normal University. Her research interests include blockchain, game theory and combinatorial optimization.
\end{IEEEbiography}
\begin{IEEEbiography}[{\includegraphics[width=1in,height=1.25in,clip,keepaspectratio]{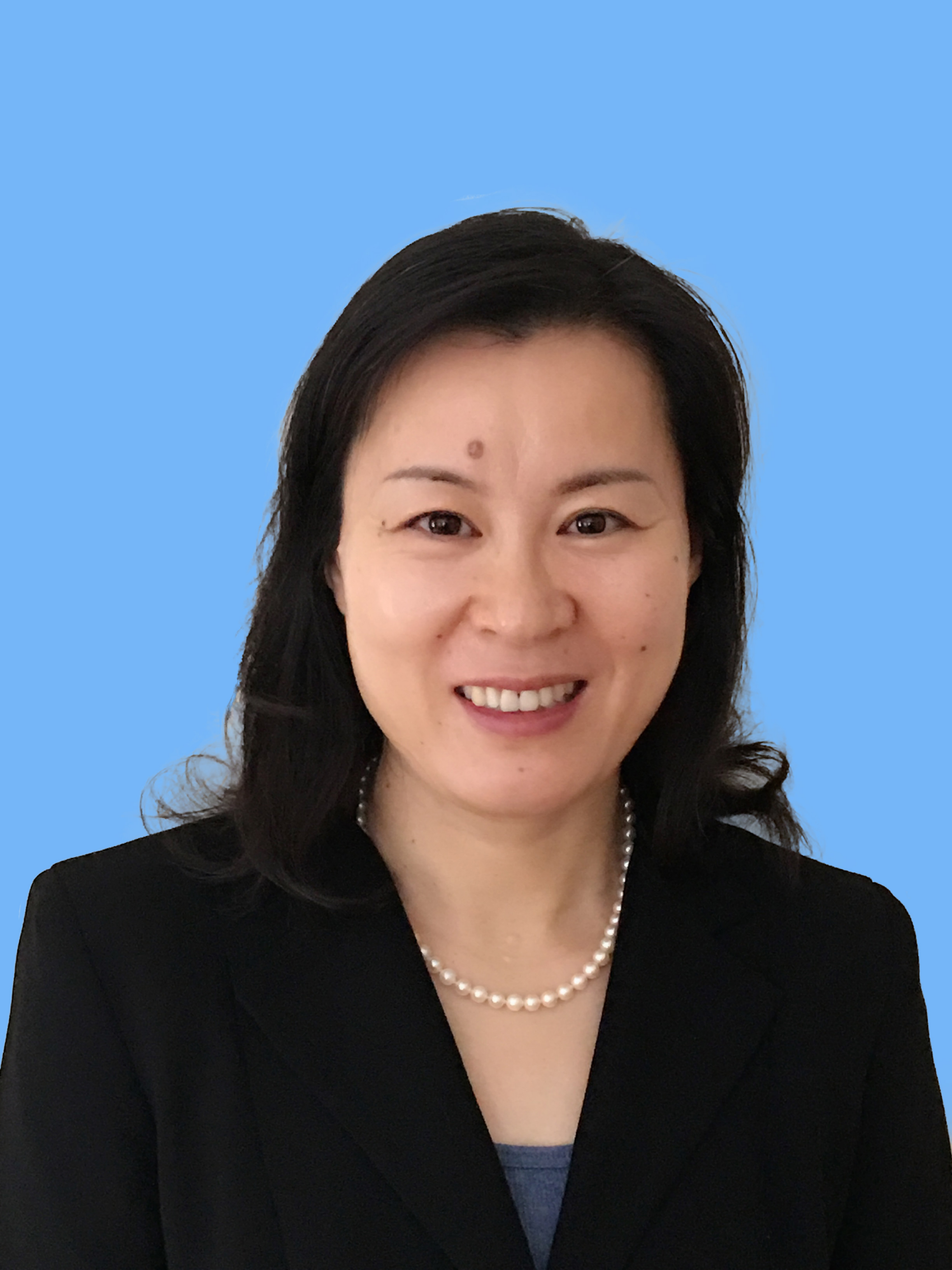}}]
	{Xiuzhen Cheng}
	received her MS and PhD degrees in computer science from University of Minnesota, Twin Cities, in 2000 and 2002, respectively. She was a faculty member at the Department of Computer Science, George Washington University,  from 2002-2020. Currently she is a professor of computer science at Shandong University, Qingdao, China. Her research focuses on blockchain computing, security and privacy, and Internet of Things. She is a Fellow of IEEE, a Fellow of CSEE, and a Fellow of AAIA.	
\end{IEEEbiography}
\begin{IEEEbiography}[{\includegraphics[width=1in,height=1.25in,clip,keepaspectratio]{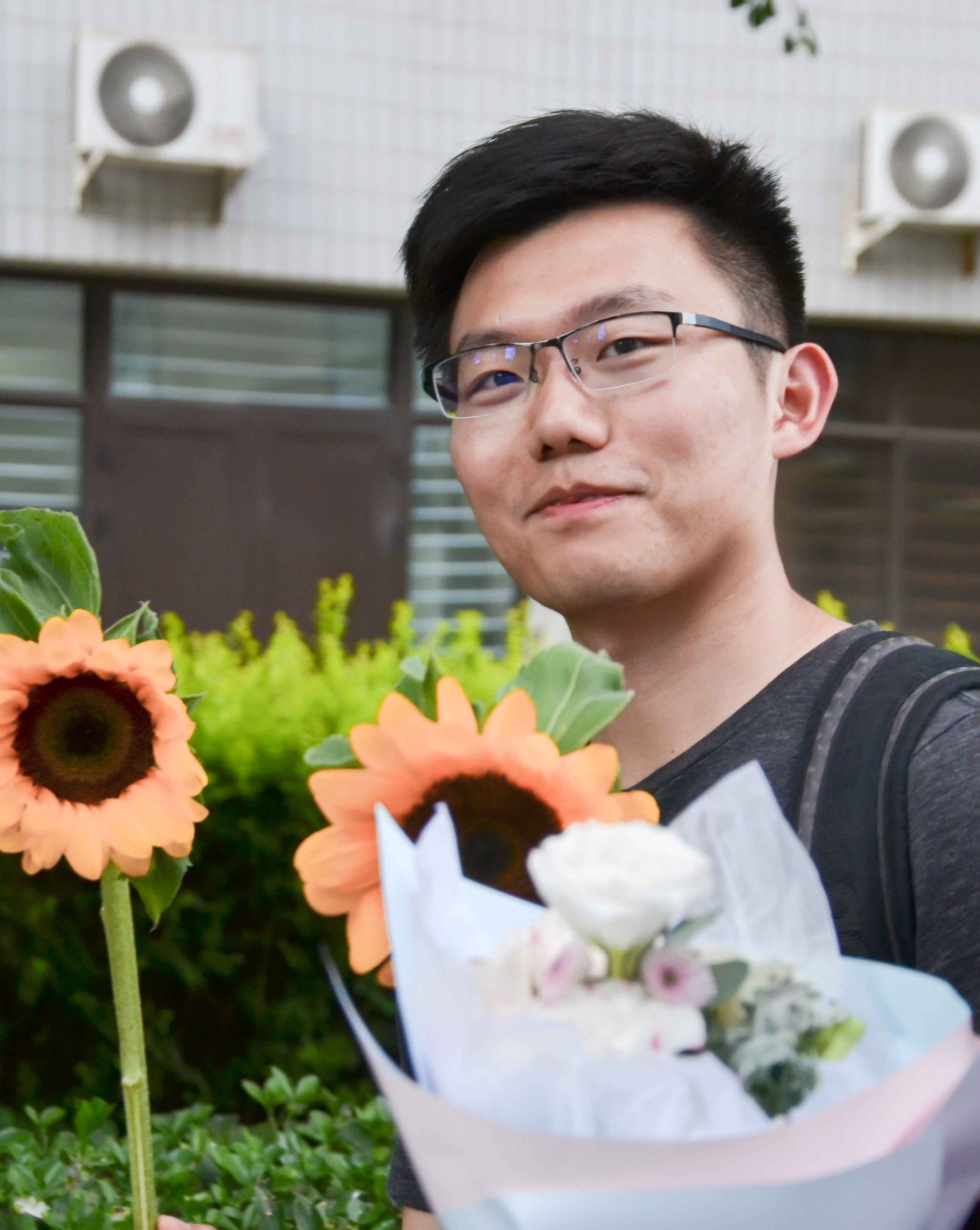}}]
	{Minghui Xu}
	is an Assistant Professor at Shandong University who received his PhD in Computer Science from The George Washington University in 2021 and his Bachelor's degree in Physics from Beijing Normal University in 2018. His research interests include blockchain, distributed computing, and cryptography.
\end{IEEEbiography}
\end{document}